\documentclass[12pt]{elsarticle}
\usepackage[utf8]{inputenc}
\usepackage{multirow}
\usepackage[normalem]{ulem}
\usepackage{url}
\usepackage{color}
\usepackage{lineno}
\usepackage{graphicx}

\let\citeauthor\relax
\let\citeyear\relax
\expandafter\let\csname ver@natbib.sty\endcsname\relax
\makeatletter
\let\c@author\relax
\makeatother
\usepackage[maxbibnames=99]{biblatex}
\journal{IET Software}

\begin{document}
\begin{frontmatter}
\title{Retrieving and mining professional experience of software practice from grey literature: an exploratory review}
\author[label1]{Austen Rainer}
\author[label2]{Ashley Williams}
\author[label1]{Vahid Garousi}
\author[label3]{Michael Felderer}

\address[label1]{School of Electronics, Electrical Engineering and Computer Science, Queen's University Belfast, Northern Ireland, U.K.}
\address[label2]{Department of Computing and Mathematics, Manchester Metropolitan University, Manchester, U.K.}
\address[label3]{Department of Computer Science, University of Innsbruck, Innsbruck, Austria; and Department of Software Engineering, Blekinge Institute of Technology, Karlskrona, Sweden}

\date{September 2020}
\begin{abstract}
\textbf{Background}: Retrieving and mining practitioners' self--reports of their professional experience of software practice could provide valuable evidence for research. We are, however, unaware of any existing reviews of research conducted in this area. \textbf{Objective}: To review and classify previous research, and to identify insights into the challenges research confronts when retrieving and mining practitioners' self-reports of their experience of software practice. \textbf{Method}: We conduct an exploratory review to identify and classify 42 articles. We analyse a selection of those articles for insights on challenges to mining professional experience. \textbf{Results}: We identify only one directly relevant article. Even then this article concerns the software professional's emotional experiences rather than the professional's reporting of behaviour and events occurring during software practice. We discuss challenges concerning: the prevalence of professional experience; definitions, models and theories; the sparseness of data; units of discourse analysis; annotator agreement; evaluation of the performance of algorithms; and the lack of replications. \textbf{Conclusion}: No directly relevant prior research appears to have been conducted in this area. \textcolor{black}{We discuss the value of reporting negative results in secondary studies.} There are a range of research opportunities but also considerable challenges. We formulate a set of guiding questions for further research in this area.
\end{abstract}
\end{frontmatter}
% \linenumbers

\section{Introduction}
\label{section:introduction}

\subsection{Motivation}
% Software practitioners form beliefs about software practice from their direct, professional experience of that practice \cite{rainer2003persuading, devanbu2016belief}. Software practitioners also form beliefs of practice indirectly from the information shared by other professionals about their respective experiences \cite{rainer2003persuading, devanbu2016belief}. 

\textcolor{black}{Software practitioners form beliefs about software practice directly from their professional experience of that practice and indirectly from the information shared by other professionals about their respective experiences \cite{rainer2003persuading, devanbu2016belief}.}

Social media provides a platform for the \textit{social programmer} \cite{storey2014r} to share their experiences of software practice online. Retrieving and mining practitioners' experience from social media therefore provides the opportunity for research to automatically and non--invasively collect evidence from practitioners about software practice. This evidence can complement evidence collected through the more traditional methods of data collection, such as interviews and surveys. More generally, retrieving and mining professional experience would contribute to the aims of Evidence Based Software Engineering (EBSE), i.e., to integrate best evidence from research with \textit{practical experience} and human values \cite{dyba2005evidence}.

Retrieving practitioners' professional experience from online, often communicated through \textit{high--quality} grey literature, is challenging \cite{rainer2019usingblogs}. There is a very large, and exponentially increasing, volume of online articles; there is considerable `noise' in the data; and experience is not articulated in text in any standard way that can be effectively searched by keyword--based search engines. And once retrieved, documents then need to be mined, e.g., to identify and extract the segments of text that represented the experience.  

Soldani \textit{et al}. \cite{soldani2018pains} and Garousi and M{\"a}ntyl{\"a} \cite{garousi2016and} both provide examples of researchers using the professional experience of the author/s of online grey literature as an inclusion/rejection criterion in their systematic grey literature reviews. In both studies, the authors manually retrieved and analysed online articles. Rainer \cite{rainer2017using} demonstrates a method for extracting arguments, experience and explanation from blog articles. As with Soldani \textit{et al}., and Garousi and M{\"a}ntyl{\"a}, Rainer's method is currently applied manually.
% (In section \ref{section:scoping-the-problem} of the current paper, we use examples from Rainer's paper to illustrate some of the challenges with mining the reporting of professional experience of software practice.)

Given the preceding discussion, we are interested in the (semi) automatic retrieval and mining of self--reports of professional experience of software practice. Ideally, we want to select articles based on the author's self--reporting of professional experience of software practice, and exclude articles that report an individual's non--professional experience or an individual's non--software professional experience.  We are unaware of any existing literature reviews of research into the (semi) automatic retrieval and mining of self--reports of professional experience of software practice. We therefore conduct and report what we believe to be the first review in this area. We position our review as an \textit{exploratory review} because, as we will show, there is very little -- \textit{if any} -- prior work to form the basis for a Systematic Mapping Study or a Systematic Literature Review.

% The preceding context motivates the following research questions:
% \begin{enumerate}
%     \item To what degree has the online retrieval of software--related professional experience, and the subsequent mining of that experience, been studied in prior research?
%     \item \textcolor{blue}{How mature are technologies (distinct from tools) for the online retrieval of software--related professional experience, and the subsequent mining of that experience?}\footnote{This RQ is unlikely to be investigated.}
% \end{enumerate}

\subsection{Aims, objectives and contributions}
The overall aim of this paper is to review empirical studies that investigate how to retrieve and mine from online articles the self--reporting of professional experience of software practice.  Our aim is translated into two objectives: 1) to classify the articles in our dataset so as to better understand prior work in this area; and 2) to identify challenges and issues with the retrieval and mining of professional experience of software practice.

The paper makes four contributions:
\begin{enumerate}
    \item We report a \textit{negative} result for previous studies i.e., our review identifies only \textit{one} directly relevant article and this single article concerns the software professional's emotional experiences rather than the software professional's reporting of behaviour and events occurring during software practice. The implication of our negative result is that -- to the best of our knowledge -- no prior research (primary or secondary) has been conducted in this area. A study that reports a negative result may raise concerns about the rigour of the enquiry (e.g., was the review conducted properly?) and about the value of the enquiry (e.g., why conduct a review when there is little to review?). We are not aware of any prior review of the kind we report here, and it is only with the conduct of this review that we discover there is, or was, little prior work to review. Phrased another way, there was no prior indication that there would be no prior research. An alternative strategy might be (have been) to relax the search and selection criteria so as to include more studies for review. But by relaxing our search strategy we change the objectives of the review and, by so doing, we fail to report a negative result and potentially contribute to publication bias. \textcolor{black}{We discuss this issue in more detail in Section \ref{section:discussion}.}
    
    \item A classification of studies related to the retrieval and mining of experience \textit{in general}. Given our first contribution, this second contribution must necessarily be more general. 
    
    \item The identification of several challenges and issues that researchers will need to confront in the retrieval and mining of professional experience of software practice. 
    
    \item A simple set of guiding questions for primary studies, to support future researchers in this area.
\end{enumerate}

\subsection{Structure of the paper}

The remainder of the paper is structured as follows. Section \ref{section:scoping-the-problem} explains the scope of our work, presenting and discussing illustrative examples of the problem being tackled, and discussing working definitions. Section \ref{section-related-work} discusses prior work. Section \ref{section:method} explains the method we use to conduct our exploratory review. Section \ref{section:classification} presents and discusses our classification of reviewed articles. Section \ref{section:analysis} then analyses several articles in more depth. Finally, Section \ref{section:discussion} reviews the objectives of the paper, \textcolor{black}{discusses the value of negative results in secondary studies}, offers guiding questions for conducting primary studies in this area, identifies threats to the validity of this study and opportunities for further research, and offers a brief conclusion.
\section{\textcolor{black}{Professional experience of software practice}}
\label{section:scoping-the-problem}

% To help scope and explain the problem we tackle, we present and discuss several illustrative examples in section \ref{subsection:illustrative-example} and then provide working definitions in section \ref{subsection:workingDefinitions}.

In this section, we present and discuss several contrasting examples of the kinds of experience we are seeking (or not seeking) to retrieve and then mine. We then provide working definitions.

% \subsection{Illustrative examples}
% \label{subsection:illustrative-example}

Table \ref{table:illustrativeExamplesExperience} and Table \ref{table:illustrativeExamplesNonexperience} present examples to illustrate the problem we are investigating. All of the examples in the two tables are taken from a single blog article \cite{spolsky2006}, \textit{Language Wars}, written by Joel Spolsky. We select these examples because they have been studied more formally in our prior work \cite{rainer2017using}. The exploratory review we report in the current paper is focused on identifying \textit{primary studies} that have: a) sought to retrieve and mine blog articles for the kinds of example presented in Table \ref{table:illustrativeExamplesExperience}, whilst b) avoiding the kinds of example presented in Table \ref{table:illustrativeExamplesNonexperience}. In other words, our exploratory review is not directly investigating the challenges arising from the examples in the two tables, but instead investigating whether and how \textit{previous work} has investigated these kinds of example, and the challenges that have arisen during those investigations.

The examples in both tables comprise fragments of verbatim text from Spolsky's blog article.
% Preceding the examples in Table \ref{table:illustrativeExamplesExperience} is a quote, from the opening paragraph of the blog article, that provides context for the examples presented in both tables.

For the first example in Table \ref{table:illustrativeExamplesExperience}, the example comprises two fragments of text that are located at different places in the source blog article. For transparency, each text fragment is uniquely identified, e.g., E1.1 and E1.2. For all examples in both tables, URLs are indicated with underlined text together with additional comments in squared parentheses. 

Example E1 describes Spolsky's experience of a group of interns working in his company. As we discuss later in this paper, story is a common approach to articulating experience. Spolsky's description in example E1 may be understood as a story: there are actors (e.g., the interns), a time and location (e.g., last summer, at the company), decisions (e.g., what language to use), actions (e.g., build Copilot using C\# and ASP.NET), outcomes (e.g., we didn't get into trouble later) and benefits (e.g., not one minute was spent on a fruitless debate). There is even the suggestion of a hero: an intern with ``enough experience'' to do a ``beautiful job architecting the ASP.NET code exactly the right way to begin with so we didn't get into trouble later''. Examples of this kind might be relevant to software engineering researchers interested in aspects of the software process (or educators interested in internships).

Example E2 describes the technology that Spolsky's company used or has developed, and how that technology has changed over time. This example might be relevant to researchers interested in the prevalence of technology--use by practitioners, or in reasons for companies to develop their own technologies rather than use existing technologies, or in how technology--use changes over time.

Examples E3 and E4 describe situations where Spolsky recognises external events and outcomes (e.g., events relating to \textit{37 Signals}) and provides his beliefs about those events and outcomes. 

Example E5 does not directly describe actual behaviours and experience but instead provides an indication of Spolsky's accumulated experience. This example is useful to researchers interested in establishing the credibility of a practitioner, e.g., as an expert.

Examples E6 and E7 are both `negative examples' in the sense that they both present arguments about which technologies one should use, but do not describe experience. Example E6 refers to evidence but does not directly present any evidence. Instead, the example implicitly relies on the \textit{reader's} experience of software practice. Example E7 refers to knowledge, which again implies expertise and experience of using the technologies.

Example E8 presents another `negative example' concerning an action taken by Spolsky, i.e., Spolsky wrote an article. While the article itself may be about software practice, the \textit{action} Spolsky describes is not itself about software practice.
% The \textit{Lord Palmerston on Programming} article might include content describing Spolksy's experience of software practice.

Notice that in several examples the presence of URL links would, in principle, allow the researcher to corroborate Spolsky's claims using independent information. As one example, the first URL link in example E1.1 directs to a website (i.e., \url{https://www.projectaardvark.com/}) that describes the experience of the interns. As a second example, example E8 provides a link to an article that might contain a description of experience.

\begin{table}[ht]
\centering
\small
\caption{Illustrative examples of experience \label{table:illustrativeExamplesExperience}}
% \processtable{Illustrative examples of experience \label{table:illustrativeExamplesExperience}}
% {\begin{tabular*}{20pc}{@{\extracolsep{\fill}}p{0.6cm} p{7.4cm}@{}}
\scalebox{0.85}{\begin{tabular*}{35pc}{@{}p{0.6cm} p{13cm}}
% {\begin{tabular*}{l l}
% \toprule
\textbf{\#} & \textbf{Verbatim quote for example}\\
\hline
\hline
% \midrule
% \multicolumn{2}{l}{\textit{Context}}\\
% C1 & ''\dots''\\
% \multicolumn{2}{l}{}\\
\multicolumn{2}{l}{\textit{Example E1: Description of behaviour}}\\
E1.1 & ``Last summer when we had a \underline{group of interns} [URL in original] build \underline{Copilot} [URL in original], we had to decide what language to use for new code\dots\\
E1.2 & \dots What we ended up doing with the interns was just telling them that they had to build it in C\# and ASP.NET. In particular, one intern, who wrote the website part of Copilot, had enough experience with ASP.NET to know what things to avoid (like viewstate) and knew to avoid the gotchas that make it impossible to have two <forms> in one page, etc. etc., so he did a beautiful job architecting the ASP.NET code exactly the right way to begin with so we didn't get into trouble later. And the benefit was that not one minute was spent debating the merits of programming languages, a fruitless debate if I've ever seen one.''\\
\multicolumn{2}{l}{}\\
\multicolumn{2}{l}{\textit{Example E2: Description of technology used and developed}}\\
E2 & ``Finally -- as to what we use -- Copilot is C\# and ASP.Net, as I mentioned, although the Windows client is written in C++. Our older in--house code is VBScript and our newer in-house code is C\#. FogBugz is written in Wasabi, a very advanced, functional--programming dialect of Basic with closures and lambdas and Rails--like active records that can be compiled down to VBScript, JavaScript, PHP4 or PHP5. Wasabi is a private, in--house language written by one of our best developers that is optimized specifically for developing FogBugz; the Wasabi compiler itself is written in C\#.''\\
\multicolumn{2}{l}{}\\
\multicolumn{2}{l}{\textit{Example E3: Observation of external event}}\\
% \hline
E3 & ``\dots yes I've heard of 37 Signals and they're making lovely Ruby on Rails apps, and making lots of money, but that's not a \textit{safe} choice for at least another year or six. I for one am scared of Ruby because (1) it displays a stunning antipathy towards Unicode and (2) it's known to be slow\dots'' (emphasis in original)\\
\multicolumn{2}{l}{}\\
\multicolumn{2}{l}{\textit{Example E4: Observation of external event}}\\
% \hline
E4 & ``Oh and I know \underline{Paul told you} [URL in original] that he made his app in Lisp and then he made millions of dollars because he made his app in Lisp, but honestly only \underline{two people} [URL in original] ever believed him and, a complete rewrite later, they won't make \textit{that} mistake again\dots'' (emphasis in original)\\
\multicolumn{2}{l}{}\\
\multicolumn{2}{l}{\textit{Example E5: Indication of author's accumulated experience}}\\
% \hline
E5 & ``\textbf{About the Author}: I'm your host, Joel Spolsky, a software developer in New York City. Since 2000, I've been writing about software development, management, business, and the Internet on this site. For my day job, I run \underline{Fog Creek Software} [URL in original], makers of \underline{FogBugz} [URL in original] --- the smart bug tracking software with the stupid name, and \underline{Fog Creek Copilot} [URL in original] --- the easiest way to provide remote tech support over the Internet, with nothing to install or configure.'' (emphasis in original)\\
% \botrule
\end{tabular*}}{}
\end{table}

\begin{table}[ht]
\centering
\small
\caption{Illustrative `negative examples', i.e., of non--experience\label{table:illustrativeExamplesNonexperience}}
% {\begin{tabular*}{20pc}{@{\extracolsep{\fill}}p{0.6cm} p{7.4cm}@{}}
{\begin{tabular*}{35pc}{@{}p{0.6cm} p{13.5cm}}
% \toprule
\textbf{\#} & \textbf{Verbatim quote for example}\\
% \midrule
\hline
\hline
\multicolumn{2}{l}{\textit{Example E6: An argument \textit{about} evidence}}\\
E6 & ``The safe answer, for the Big Enterprisy Thing where you have no interest in being on the cutting edge, is C\#, Java, PHP, or Python, since there's so much evidence that when it comes right down to it zillions of people are building huge business--critical things in those languages and while they may have problems, they're not life--threatening problems.''\\
\multicolumn{2}{l}{}\\
\multicolumn{2}{l}{\textit{Example E7: An argument \textit{about} experience}}\\
E7 & ``How do you decide between C\#, Java, PHP, and Python? The only real difference is which \textit{one} you know better. If you have a serious Java guru on your team who has build several large systems successfully with Java, you're going to be a hell of a lot more successful with Java than with C\#, not because Java is a better language (it's not, but the differences are too minor to matter) but because he knows it better. Etc'' (emphasis in original)\\
\multicolumn{2}{l}{}\\
\multicolumn{2}{l}{\textit{Example E8: Something}}\\
% \hline
E8 & ``A while ago I wrote an article called \underline{Lord Palmerston on Programming} [URL in original] in which I claimed that some of these programming worlds, like .NET and Java, were so huge and complicated that you never could really tell if they were going to be good enough for your needs until it was too late.'' (emphasis in original)\\
% \botrule
\end{tabular*}}{}
\end{table}

% \subsection{Working definitions}
% \label{subsection:workingDefinitions}

Precisely defining the relevant concepts for this review is not straightforward, e.g., because the field is still emerging. Consequently, we present and use working definitions.
% , and then briefly review these definitions in Section \ref{section:discussion}.

We recognise the distinction between an experience itself and information shared to convey that experience. Our focus here is on information shared to convey that experience, as we cannot gain direct access to the experience itself. The examples in Table \ref{table:illustrativeExamplesExperience} all present information to convey experience. For simplicity in writing, we refer to that information--sharing as the reporting of professional experience. The examples in Table \ref{table:illustrativeExamplesExperience} suggest at least the following types of \textit{reporting of professional experience}:

\begin{description}

\item \textit{Type I}: A person's description of their own behaviour (e.g., decisions, actions) as they interact with other individuals within a professional situation, e.g., a nurse's description of her or his behaviour within a particular hospital situation. Example E1 and E2 both contain elements of this type of reporting: E1 refers to ``what we ended up doing'', and E2 refers to ``what we use''.

\item \textit{Type II}: A person's description of their direct observations of others' behaviour. Examples E1, E3 and E4 present elements of this type: E1 refers to the behaviour of interns, E3 refers to events relating to the company \textit{37 Signals}, and E4 refers to ``Paul's'' behaviour.

\item \textit{Type III}: A description of a person's accumulated experience. Example E5 presents elements of this type. This accumulated experience might then be used to assess the credibility of the practitioner, and from that the credibility of the practitioner's claims and arguments, such as those presented in examples E6 and E7.

\end{description}

% We define the \textit{reporting of professional experience} as a person's description of their own behaviour (e.g., decisions, actions) as they interact with other individuals within a professional situation (e.g., a nurse's description of her or his behaviour within a particular hospital situation) or a person's description of their direct observations of others' behaviour (e.g., a nurse's description of her or his direct observation of others interacting in particular hospital situation).

% The reporting of professional experience may, from an alternative perspective, be understood as a kind of witness statement or as a kind of expert testimony.

The reporting of the professional experience of a \textit{software practitioner} would describe her or his behaviour in relation to the domain of software engineering: designing, testing, deploying and maintaining software--based systems. It follows that there are other \textit{domains} of experience. Whilst we recognise that there are different domains of non--professional and professional experience, we consider it premature to \textit{a priori} specify or enumerate those domains. We identify some of the domains as part of our review.

We distinguish in principle the reporting of a professional \textit{experience} of actual behaviour and events (e.g., a software practitioner describes in writing her experience working on a software project) from the articulation of \textit{beliefs} based on or inferred from that experience (e.g., that practitioner's opinions about some aspect of that project). Examples E1 and E6 provide illustrative examples of this distinction. One reason we make this distinction is to help separate beliefs based on one's direct experience from beliefs formed from information provided by others. In other words, we seek to confine a witness's statement to a description of what occurred, and distinguish the witness statement from expert testimony. In practice, it may be very challenging to confine such witness statements and to distinguish them from expert testimony.

Just as there are challenges isolating description from opinion, so there will likely be challenges separating factual information from other kinds of information, such as evaluative and emotional information, e.g., examples E3, where Spolsky writes of being ``scared''. These are some of the challenges that the successful retrieval and mining of professional experience will need to tackle.

Finally, whilst there is no standard way for reporting professional experience, we observed in example E1 --- and discuss later in the paper --- that \textit{story} appears to be a common approach used by people to report their own and others' behaviour. Again, difficulties arise between the use of a factual or partially factual story to describe an experience from the past, in contrast to an abstract story describing a desired future state (e.g., \textit{User stories}), or in contrast to fictional stories (e.g., \cite{rainer2020anaya}) used for entertainment or to convey moral value or other meanings. A further challenge is that, for reasons of confidentiality, commercial sensitivity or diplomacy etc., a contrived story may be used to illustrate a real situation.
\section{Related work}
\label{section-related-work}

\subsection{The value of practitioner experience}

Evidence Based Software Engineering (EBSE) seeks to integrate best evidence from research with practical experience and human values \cite{dyba2005evidence}. Devanbu \textit{et al.} \cite{devanbu2016belief} and Rainer \textit{et al.} \cite{rainer2003persuading}, amongst others, have observed that software practitioners value their own experience, and that of their peers, over empirical evidence, placing most value on local expertise.

Storey \textit{et al}. \cite{storey2014r} discuss the emergence of the \textit{social programmer}, as social media has dramatically changed the way that software practitioners share information. There are large amounts of online articles, as people share information (though not necessarily about professional issues, or about their experience of practice).
% \textcolor{blue}{\sout{For example, Lakshmanan and Oberhofer \cite{lakshmanan2010knowledge} report that Bansal \textit{et al}.'s \textit{BlogScope} \cite{bansal2007seeking} ``\dots currently tracks more than 36.88 million blogs with 837.39 million posts in the blogosphere\dots [fetching on average] 14,000 new documents per hour.''. Bansal \textit{et al}.'s \textit{BlogScope} was developed over a decade ago, is no longer active, and was not specifically focused on software practice.}}
In terms of software practice, Choi \cite{kilimchoi18} maintains a reference list of 650 blogs, and Soldani \textit{et al}. \cite{soldani2018pains} observe a ``\dots massive proliferation of grey literature [on microservices], with more than 10,000 articles on disparate sub--topics\dots''.

\textcolor{black}{Overall, we conjecture that, in due course, experience mining might help research to achieve EBSE, i.e., by integrating professional experience of software practice with best evidence from software engineering research with human values.}

\subsection{Grey literature in software engineering research}
There is emerging interest in incorporating grey literature into software engineering research as such literature provides insight into industry practice. Garousi \textit{et al.} \cite{garousi2016need} advocate the use of Multi-Vocal Literature Reviews (MLRs) to synthesise peer--reviewed academic literature with grey literature. Such syntheses can help to address the disconnect between the state--of--art and the state--of--practice. (Garousi \textit{et al.} have published guidelines for conducting MLRs \cite{garousi2017guidelines}.) There is the need to automate aspects of the review process to handle the large volume of articles being considered. Rainer \cite{rainer2017using} presents a methodology for analysing one type of grey literature, the practitioner--written blog article, for reports of the software practitioner's experience and for the beliefs that practitioners infer from those experiences. Rainer recognised the need to automate the methodology so that it could be scaled to the large volume of online articles produced by software practitioners.
% Experience mining, together with argumentation mining, offer two fields of research to automate such analyses.

% The following paragraphs are commented out to save space for EASE'19

\textcolor{black}{A comparable situation is found in the health sector: Adams \textit{et al.} found that many evaluations of public health interventions ``\dots may be predominantly, or only, held in grey literature and grey information\dots'' \cite{adams2016searching}.
For Adams \textit{et al.} there are at least three reasons why researchers may want to include grey literature in their studies. (We rephrase them here in terms of software engineering research.) First, including grey literature can reduce publication bias. Studies with null findings are less likely to be published in peer--reviewed software engineering journals and conferences, so grey literature can help to compensate for unpublished null findings. Second, grey literature can provide useful contextual information on how, why, and for whom software engineering practices and technologies are effective: context is recognised as a significant challenge in software engineering research. Third, synthesising grey literature can help software engineering researchers and practitioners understand what solutions exist for a particular software engineering problem, the full range of evaluations (if any) that have been conducted, and where further development and evaluation is needed.}

% In previous research, we have begun to develop a methodology to help researchers search for and filter grey literature, in order to identify the higher--quality grey literature for software engineering research \cite{williams_rainer}. We defined quality in terms of a document's credibility. As credibility is at least partly a subjective concept, we have  conducted a survey of software engineering researchers \cite{williams_rainer_survey} to better understand the criteria they use for assessing credibility. Respondents to the survey were drawn from the programme committees of two leading international conferences (\textit{Evaluation and Assessment in Software Engineering} and \textit{Empirical Software Engineering and Measurement}). The survey found that the reporting of personal experience was an important criterion to determine an article's credibility. 

% Rainer \textit{et al.} \cite{rainer2003persuading} and Devanbu \textit{et al.} \cite{devanbu2016belief} have both argued that whilst researchers highly value empirical data, practitioners form opinions primarily based on their personal experience. Therefore, detecting experience within grey literature is an important step in the (semi--)automated assessment of credibility and, by inference (according to our definition), of the quality of practitioner--written articles.

\textcolor{black}{Overall, we conjecture that, in due course, experience mining might help research automate MLRs and grey literature reviews (GLRs).}

\subsection{Mining opinions, arguments and experiences}
%\textcolor{magenta}{This section requires some work. I currently think we can talk generally, in terms of people like Bex and Twining; and then cite Lippi and Whoever as our main source.}

Cabrio and Villata \cite{cabrio2018five} present a five--year review (c2014 --- c2018) of argumentation mining, complementing the previous reviews of Lippi and Torroni \cite{lippi2016argumentation} and Peldszus and Stede \cite{peldszus2013argument}. (Mochales and Moens \cite{mochales2011argumentation} provide an earlier review restricted to argumentation mining in relation to law.)

Cabrio and Villata \cite{cabrio2018five} distinguish \textit{opinion mining}, which they consider focuses on understanding \textit{what} users think about a topic, from \textit{argument mining}, which focuses on \textit{why} users have a certain opinion.
% (This could be re-conceived as the distinction between what users' \textit{believe}\cite{devanbu2016belief, rainer2003persuading}, with the prospect of a research field of \textit{belief mining}, and the arguments to justify their \textit{beliefs}.)\\
To these we may add \textit{experience mining}, which focuses on a person's experiences, distinguished from their opinion and argument.
% (and which we explore later in this paper).
% , and \textit{citation mining}, which focuses on mining (explicit) references to other sources of information, such as the white and grey literature.

Cabrio and Villata \cite{cabrio2018five} distinguish the identification of an argument or arguments within a text, from predicting the relations between the extracted arguments. They recognise that argumentation mining requires high--quality annotated corpora, that there is the additional challenge in argumentation mining of establishing the factuality of information in an argument; and that there is the significant challenge of aligning the heterogeneity of datasets and approaches used by researchers, partly explained by the immaturity of the field: each dataset has been annotated on the basis of different definitions of argument components and relations, thus preventing the possibility of a straightforward alignment of these datasets.

% Lippi and Torroni \cite{lippi2016argumentation} structure their review of previous argumentation mining research around their argumentation pipeline. A version of the pipeline (based on Lippi and Torroni's \cite{lippi2016margot}) is presented in Figure \ref{figure:argumentation-pipeline}, with brief explanations in Table \ref{table:argumentation-pipeline-explanation}. The figure has been modified to include our continuing research on the search and selection of grey literature \cite{rainerIST18shortpaper}. 

\begin{figure*}[ht]
    \centering
    \includegraphics[width=0.9\textwidth]{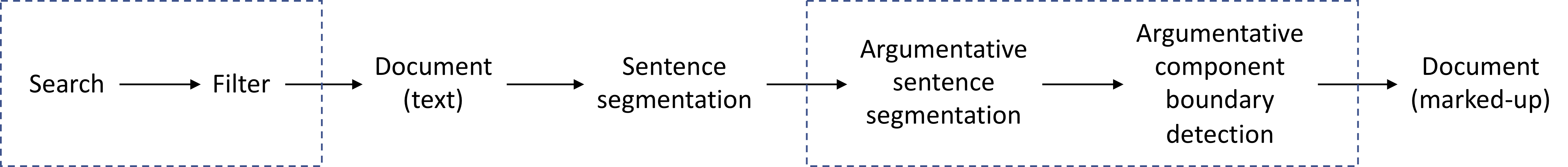}
    \caption{Argumentation pipeline, modified from \cite{lippi2016margot}}
    \label{figure:argumentation-pipeline}
\end{figure*}

Lippi and Torroni \cite{lippi2016argumentation} structure their review around their argumentation pipeline. A version of the pipeline (based on Lippi and Torroni's \cite{lippi2016margot}) is presented in Figure \ref{figure:argumentation-pipeline}, modified to include our continuing research on the search and selection of grey literature \cite{rainerIST18shortpaper}. Lippi and Torroni's pipeline uses the sentence as the basic unit of discourse. As we show later in this paper, previous research on experience mining has used a variety of units of discourse, for which the sentence is one.

Lippi and Torroni \cite{lippi2016margot} use an IBM--developed annotated corpus\footnote{\url{https://www.research.ibm.com/haifa/dept/vst/mlta_data.shtml}} to evaluate their MARGOT (Mining ARGuments frOm Text) software system, the first publicly accessible web server for argumentation mining. The corpus was developed by IBM as part of its Debater project. The annotation guidelines for the IBM corpus distinguish three types of \textit{evidence} for arguments: \textit{study} (e.g. results of quantitative analyses), \textit{expert} (e.g. testimony by someone/s with known expertise on, or authority in, some topic) and \textit{anecdotal} (a description of specific event instances or concrete examples). It is the third type of evidence that is the particular focus of the current paper (though the second type is relevant too).
% (Lippi and Torroni \cite{lippi2016argumentation} also review argumentation mining research, structuring that review around an argumentation pipeline.)

Other research on argumentation defines experience in terms of detecting events and stories. Bex \cite{bex2013values} defines stories as ``\dots a coherent sequence of events, often involving subjects, objects, outcomes and other attributes'', and Twining \cite{twining1994rethinking} defines a story as ``\dots a narrative of particular events arranged in a time sequence and forming a meaningful totality.''

\textcolor{black}{Overall, we conjecture that experience mining might effectively complement the existing sub--fields of text mining.}

\subsection{Tool support for systematic reviews}

Software engineering research has investigated how to reduce the time and effort for systematic reviews. For example, Marshall \textit{et al}. \cite{marshall2013tools, marshall2014tools} have reviewed tools to support systematic reviews (SR) in software engineering. These tools support a number of different steps in the SR process. None of the tools appear to consider the mining of professional experience.

Unsurprisingly, evidence based medicine has considered some of the issues we raise in this paper. Tsafnat \textit{ et al}. \cite{Tsafnat2014} report a survey of tools for automating parts of (evidence based medicine's) systemic reviews, structured in terms of the steps of a review (such as search and extraction). Their survey does not appear to address the mining of professional experience. O’Mara-Eves \textit{ et al}. \cite{OMara-Eves2015} report the systematic review of text mining for study identification. They observe an active and diverse body of research, but with almost no replication between studies or collaboration between research teams. They conclude that text mining is ready to be used in `live' systematic reviews (in medicine) for \textit{prioritising} the order in which items are reviewed; and may be used cautiously as a `second screener' to complement the human reviewer. Their focus appears to be on \textit{screening} articles.

\textcolor{black}{Overall, we conjecture that experience mining might inform the development of appropriate software tools.}

\section{Method}
\label{section:method}

In the following subsections, we first describe our process for searching and selecting articles, then explain our rationale for performing two sets of analyses, and then describe how each set of analyses was conducted.

\subsection{Searching \& selecting articles}
The second author performed 18 queries, using five search engines, to identify the initial set of candidate articles. These searches were conducted in July 2018. We followed the recommendations of Singh and Singh \cite{singh2017exploring} to tailor our search strings to the specific requirements of each search engine. Table \ref{table:search-queries} summarises the searches. Table \ref{table:filtering} summarises the steps involved in the subsequent filtering of articles.

\begin{table*}[ht]
% \centering
% \small
% \caption{Summary of searches conducted}
% \label{table:search-queries}
% \begin{tabular}{|c|p{5.5cm}|r|r|r|r|r|}
% \fwprocesstable{Summary of searches conducted\label{table:search-queries}}
\caption{Summary of searches conducted\label{table:search-queries}}
\scalebox{0.85}{\begin{tabular*}{39pc}{ c p{6cm} r r r r r}
% {\begin{tabular*}{}{c p{6cm} r r r r r }
% \toprule
% \hline
\multirow{3}{*}{\textbf{\#}} & \multirow{3}{*}{\textbf{Search Strings}}                   & \multicolumn{5}{c}{\textbf{Results per search engine}}\\
% \cline{3-7} 
& & \textbf{Google} & \textbf{IEEE } & \textbf{} & \textbf{Science} & \textbf{} \\
& & \textbf{Scholar} & \textbf{Xplore} & \textbf{ACM} & \textbf{ Direct} & \textbf{Scopus} \\
% \midrule
\hline
\hline
\textbf{1}                 & ``experience mining''                                        & 664                     & 8                    & 9            & 40                      & 105             \\
% \hline
\textbf{2}                 & ``experience indicators''                                    & 1,850                    & 2                    & 534          & 160                     & 82              \\
% \hline
\textbf{3}                 & indicators AND ``personal experience''                       & 132,000                 & 3                    & 25           & 9,977                    & 4,148            \\
% \hline
\textbf{4}                 & indicators AND ``professional experience''                   & 43,500                  & 1                    & 6            & 2,582                    & 549             \\
% \hline
\textbf{5}                 & indicators AND experience                                  & 3,680,000               & 1,199                 & 5,176         & 407,703                  & 291,425          \\
% \hline
\textbf{6}                 & allintitle: indicators personal experience                 & 2                       & 81                   & 0            & 0                       & 1               \\
% \hline
\textbf{7}                 & allintitle: indicators personal experience online articles & 0                       & 27                   & 0            & 0                       & 0               \\
% \hline
\textbf{8}                 & allintitle: indicators experience web                      & 3                       & 8,675                 & 1            & 0                       & 0               \\
% \hline
\textbf{9}                 & allintitle: ``experience indicators''                        & 38                      & 0                    & 0            & 0                       & 7               \\
% \hline
\textbf{10}                & allintitle: detecting experience                           & 266                     & 3,118                 & 65           & 87                      & 162             \\
% \hline
\textbf{11}                & allintitle: ``detecting experience''                         & 2                       & 0                    & 4            & 0                       & 2               \\ 
% \hline
\textbf{12}                & allintitle: ``detecting'' AND ``personal experience''          & 4                       & 0                    & 0            & 0                       & 1               \\
% \hline
\textbf{13}                & allintitle: ``detecting'' AND ``professional experience''      & 0                       & 0                    & 0            & 0                       & 0               \\
% \hline
\textbf{14}                & ``detecting professional experience''                        & 0                       & 0                    & 0            & 0                       & 0               \\
% \hline
\textbf{15}                & ``detecting personal experience''                            & 32                      & 0                    & 0            & 3                       & 0               \\
% \hline
\textbf{16}                & ``detecting experience''                                     & 67                      & 0                    & 30           & 62                      & 26              \\
% \hline
\textbf{17}                & ``story mining”                                             & 209                     & 2                    & 1            & 5                       & 12              \\
% \hline
\textbf{18}                & ``detecting stories”                                        & 63                      & 1                    & 5            & 19                      & 25              \\
% \hline
% \botrule
\end{tabular*}}{}
\end{table*}

For each search query, the first 50 results were considered for inclusion in the review. (Given that, subsequently, only 15\% of the search results were classified, and less than 1\% of the search results are directly relevant to our topic, selecting the first 50 search results was an effective, pragmatic inclusion criteria for this study.) Each of those 50 results was selected as a candidate if that result could \textit{not} be immediately rejected based on the respective article's title, e.g., an article was removed where its title was referring to mineral mining. Results that were written in a non--English language, or that were patents, were \textit{retained} at this stage, because they could not be immediately rejected based on title. The initial search process yielded 309 candidate articles.
% Removing duplicate articles reduced the number of articles to 238 articles. 

The second author then filtered the 309 articles following the steps summarised in Table \ref{table:filtering}. Following the filtering, 42 articles remain for classification.

% \begin{table}[!b]
% \processtable{Example table\label{tab1}}
% {\begin{tabular*}{20pc}{@{\extracolsep{\fill}}lll@{}}\toprule
% Column  &Column  & Column heading \\
% heading  &heading two &  three \\
% \midrule
% Row 1a  &Row 1b  &Row 1c \\
% Row 2a  &Row 2b  &Row 2c \\
% Row 3a  &Row 3b  & Row 3c \\
% Row 4a  &Row 4b  &Row 4c \\
% Row 5a  &Row 5b  &Row 5c \\
% Row 6a  & Row 6b  & Row 6c \\
% \botrule
% \end{tabular*}}{}
% \end{table}

\begin{table}[ht]
\centering
% \small
% \processtable{Summary of filtering\label{table:filtering}}
\caption{Summary of filtering\label{table:filtering}}
% \caption{Summary of filtering}
% \label{table:filtering}
% \begin{tabular}{|p{5.8cm}|p{1.5cm}|}
{\begin{tabular*}{13.5cm}{ p{10.5cm} p{1.5cm}}
% \toprule
% \hline
\textbf{Filtering step} & \textbf{Remaining articles}\\
\hline
\hline
% \midrule
\multicolumn{2}{l}{}\\
Initial results from searches & \centering{309} \tabularnewline 
% \hline
% \midrule
\multicolumn{2}{l}{}\\
Obvious duplicate articles removed ($n=71$) & \centering{238} \tabularnewline 
% \hline
% \midrule
\multicolumn{2}{l}{}\\
Review article's abstract, introduction and conclusion to remove irrelevant articles, i.e., did the article discuss detecting experience, stories or events? ($n=154$) & \centering{84} \tabularnewline
% ($n=154$) & \tabularnewline
% \hline
% \midrule
\multicolumn{2}{l}{}\\
Non--accessible articles removed ($n=21$) & \centering{63} \tabularnewline
\hspace{0.1cm} No abstract or article ($n=15$) &\\
\hspace{0.1cm} Abstract available, but not article ($n=6$) &\\
% \hline
% \midrule
\multicolumn{2}{l}{}\\
\textcolor{black}{Apply exclusion criteria, e.g., patents, similar papers, datasets only, theses \& duplicate ($n=21$)} & \centering{42} \tabularnewline
% \midrule
% \hspace{0.1cm} patents &\\
% \hspace{0.1cm} similar papers &\\
% \hspace{0.1cm} datasets only &\\
% \hspace{0.1cm} theses&\\
% \hspace{0.1cm} duplicate&\\
% Apply exclusion criteria ($n=14+?+?+2+3+1$) & \centering{43} \tabularnewline
% \hspace{0.1cm} patents ($n=2$) &\\
% \hspace{0.1cm} similar papers \textcolor{blue}{$n=13?$} &\\
% \hspace{0.1cm} datasets only \textcolor{blue}{$n=1?$} &\\
% \hspace{0.1cm} theses ($n=3$) &\\
% \hspace{0.1cm} duplicate ($n=1$) &\\
% \hline
% Remove patents ($n=2$) & 46\\
% \hline
% Remove other articles during analysis ($n=1$) & \centering{42} \tabularnewline
% \hline
% \botrule
\end{tabular*}}{}
\end{table}

\subsection{Rationale for our two sets of analyses}

\textcolor{black}{We perform two sets of analyses of the 42 articles: first, a classification of the articles; second, a comparison of a selection of articles. Our classification finds only \textit{one} article that is directly concerned with the retrieval and/or subsequent mining of the reporting of professional experience of software practice. Furthermore, this singular article concerns the software professional's \textit{emotional experiences} rather than the software professional's reporting of behaviour and events occurring during software practice. Our second set of analysis therefore compares a \textit{selection} of articles for their insights into challenges confronted by, and issues recognised by, the researchers in their respective research. We expect these challenges will recur as and when researchers investigate the retrieval and mining of professional experience of software practice.}

% Our first set of analysis provides a broad, top--down classification of the articles. Because of the limited number of directly relevant articles, we both discuss other aspects of our first classification (so as to maximise the value of that classification) and also perform our second set of analysis (again to maximise what we can learn from the dataset). Our second set of analysis compares a selection of articles for their insights into challenges confronted by, and issues recognised by, the researchers in their respective research.

% Our first set of analysis therefore allows us to position research on the retrieval and mining of professional experience of software practice within a wider `experience space'. Our second analysis allows us to identify challenges and issues that we can expect will occur as and when researchers investigate the retrieval and mining of professional experience of software practice.

\subsection{Classifying the full set of articles}

The first and second authors initially reviewed all 42 articles to better understand the full set of articles, e.g., reviewing a subset of articles and then returning to re--consider the full set of articles. (We actually reviewed a slightly larger set, 48 articles, but excluded 6 of these for reasons indicated in Table \ref{table:filtering}.) This initial review produced a rudimentary classification scheme and a preliminary classification of the articles.

The first author then developed a simple classification framework to be used to re--classify the full set of 42 articles. \textcolor{black}{All 42 articles were independently classified by the first and the second authors. As a further check, and to enable all authors to contribute to the full paper, the third and fourth authors each independently classified a quarter of the articles, and a colleague also independently classified a quarter, resulting in the majority of articles being independently reviewed.} Differences in classifications were resolved by the first author.

% This classification framework was then used independently, by each of the other co--authors of the paper, to re--classify the articles. The second author independently classified all 42 articles. The two other authors (together with an additional colleague) each independently classified a subset of 12 articles each. Differences in classifications were resolved by the first author.

\subsection{Analysing a subset of articles}

\textcolor{black}{During the development of the classification for our first set of analyses, the first and second authors developed a deeper appreciation of the full set of articles. We selected eight articles for further comparison. These articles were selected because they offered insights into the challenges confronted by, and the issues recognised by, the researchers involved in the primary studies. We do not claim that we provide a complete coverage of all challenges. Of these eight articles, seven were selected from the list of 42 articles. We identified an eighth article \cite{ceran2012hybrid} from our fuller set of articles. We included that eighth article  because the article provided a useful contrasting datapoint. We emphasise therefore that the second analyses does not analyse a subset of the 42 articles but instead analyses a selection of articles to identify and illustrate challenges. The second analyses does not claim to be exhaustive in the challenges identified.}

\section{Classification of the articles}
\label{section:classification}

% }
% % Table \ref{table:frequency-of-publication} presents a frequency analysis of the papers published each year since 2003.}

% \textcolor{blue}{Table \ref{table:classification-of-46-papers}

\textcolor{black}{Table \ref{table:classification-of-46-papers} presents our classification of the 42 papers. The table  suggests that the first papers explicitly investigating experience mining appear to be Kurashima \textit{et al}. \cite{kurashima2006mining} in 2006, followed by Inui \textit{et al}. \cite{inui2008experience} in 2008. Table \ref{table:classification-of-46-papers} includes two papers \cite{brants2003system, lee2004multilingual} that precede Kurashima \textit{et al}. \cite{kurashima2006mining} and Inui \textit{et al} \cite{inui2008experience}; a third paper \cite{qamra2006mining} was published in the same year. These three papers are concerned with story detection and event detection rather than directly with experience mining.}

\textcolor{black}{Table \ref{table:classification-of-46-papers} shows that there is only one article \cite{fontao2017facing} that is directly concerned with the retrieval and subsequent mining of the reporting of professional experience of software practice.  Even then, the article focuses more on the developers' emotional responses to situations rather than the developers recounting of their experiences.}

\textcolor{black}{Table \ref{table:classification-of-46-papers} also suggests that there are few articles on the mining of professional experience in general: there are many more articles that investigate the reporting of non--professional (personal) experience compared to professional experience. The two articles by Gon{\c{c}}alves \textit{et al}. \cite{de2011collaborative,de2010case} are the only articles in our dataset to examine professional experience. They use a tool to \textit{generate} stories from the professional participants in contrast to, for example, mining reports of professional experience from practitioners' grey literature. Gon{\c{c}}alves \textit{et al} also intend to use the generated stories to define business rules and processes, rather than as evidence for research.}

\textcolor{black}{There are a number of articles that detect experience in terms of story although, by contrast, the one article \cite{fontao2017facing} concerned with detecting the reporting of professional experience of software practice does \textit{not} detect experience in terms of story. The table identifies those articles concerned with story and organises them into those that detect professional experience and those that detect non--professional experience. Though not shown in the table, there are a number of different types of story: story as public news (e.g., a news story), story as a stream of events, folk stories, telling stories. Table \ref{table:elements-in-story-based-models-of-experience} presents examples of story based and non--story based experience.}

% \textcolor{blue}{Turning to Table \ref{table:frequency-of-publication}, the table indicates that, on
\textcolor{black}{On average, about three papers are published per year in this area, however 75\% of the papers have been published in the most recent eight years (the second half of the sixteen year period in which the set of articles have been published). This suggests a slight growth in the area.}

\begin{table}
\centering
    % \small
    % \caption{Classification of the retained articles}
    % \label{table:classification-of-46-papers}
    %\begin{tabular}{| p{0.6cm} | p{0.7cm} | p{0.7cm} | p{0.7cm} | p{0.7cm} | p{0.7cm} | p{0.7cm} | p{0.7cm} | p{0.7cm} | p{1.8cm} | p{0.7cm} | p{0.7cm}|}

    \caption{Classification of the retained articles\label{table:classification-of-46-papers}}
    \scalebox{0.70}{\begin{tabular*}{32pc}{@{\extracolsep{\fill}} c l l l l l l l l l l @{}}
    % \toprule

    % \begin{tabular}{| c |l |l |l |l |l |l |l |l |l |l |}
    % \hline
    % \textbf{Ref}	&	\textbf{Year}	&	\textbf{Nws}	&	\textbf{SP}	&	\textbf{Prof}	&	\textbf{Per}	&	\textbf{Blg}	&	\textbf{Emt}	&	\textbf{Domain}	&	\textbf{Stry}	&	\textbf{Twt}	\\
    \textbf{Ref}	&	\textbf{Year}	&	\textbf{Nw}	&	\textbf{SP}	&	\textbf{PE}	&	\textbf{Pl}	&	\textbf{Bl}	&	\textbf{Em}	&	\textbf{Domain}	&	\textbf{St}	&	\textbf{Tw}	\\
    \hline
    \hline
    % \midrule
    \multicolumn{11}{ l }{\textit{Articles about the experience of software practice}}\\
    % \hline		
    \cite{fontao2017facing} 	&	2017		&		&	\multicolumn{1}{c}{$\bullet$}	&	\multicolumn{1}{c}{$\bullet$}	&		&		&	\multicolumn{1}{c}{$\bullet$}	&	Mobile app	&		&		\\
    % \hline
    % \midrule
    \multicolumn{11}{ l }{\textit{Articles that represent  non--software professional experience in terms of story }}\\
    % \hline
    \cite{de2011collaborative} 	&	2011	&		&		&	\multicolumn{1}{c}{$\bullet$}	&		&		&		&	Processes	&	\multicolumn{1}{c}{$\bullet$}	&		\\
    \cite{de2010case} 	&	2010	&		&		&	\multicolumn{1}{c}{$\bullet$}	&	&		&		&	Processes	&	\multicolumn{1}{c}{$\bullet$}	&		\\
    
    % \hline
    % \midrule
    \multicolumn{11}{ l }{\textit{Articles that represent  non--professional experience in terms of story}}\\
    % \hline
    
    \cite{swanson2014identifying} 	&	2014	&		&		&		&	\multicolumn{1}{c}{$\bullet$}	&	\multicolumn{1}{c}{$\bullet$}	&	\multicolumn{1}{c}{$\bullet$}	&	Personal	&	\multicolumn{1}{c}{$\bullet$}	&		\\
    \cite{ceran2012asemantic} 	&	2012	&		&		&		&	\multicolumn{1}{c}{$\bullet$}	&		&		&	Extremists	&	\multicolumn{1}{c}{$\bullet$}	&		\\
    
    \cite{gruenheid2015storypivot} 	&	2015	&	\multicolumn{1}{c}{$\bullet$}	&		&		&	\multicolumn{1}{c}{$\bullet$}	&	\multicolumn{1}{c}{$\bullet$}	&		&	News	&	\multicolumn{1}{c}{$\bullet$}	&		\\
    \cite{DiCrescenzo2017hermevent}	&	2017	&	\multicolumn{1}{c}{$\bullet$}	&		&		&	\multicolumn{1}{c}{$\bullet$}	&		&		&	News	&	\multicolumn{1}{c}{$\bullet$}	&	\multicolumn{1}{c}{$\bullet$}	\\
    \cite{srijith2017sub} 	&	2017	&	\multicolumn{1}{c}{$\bullet$}	&		&		&	\multicolumn{1}{c}{$\bullet$}	&		&		&	News	&	\multicolumn{1}{c}{$\bullet$}	&	\multicolumn{1}{c}{$\bullet$}	\\
    \cite{petrovic2012using}	&	2012	&	\multicolumn{1}{c}{$\bullet$}	&		&		&	\multicolumn{1}{c}{$\bullet$}	&		&		&	News	&	\multicolumn{1}{c}{$\bullet$}	&	\multicolumn{1}{c}{$\bullet$}	\\
    \cite{petrovic2010streaming} 	&	2010	&	\multicolumn{1}{c}{$\bullet$}	&		&		&	\multicolumn{1}{c}{$\bullet$}	&		&		&	News	&	\multicolumn{1}{c}{$\bullet$}	&	\multicolumn{1}{c}{$\bullet$}	\\
    \cite{yu2018learning} 	&	2018	&	\multicolumn{1}{c}{$\bullet$}	&		&		&		&		&		&	News	&	\multicolumn{1}{c}{$\bullet$}	&		\\
    \cite{lee2006korean} 	&	2006	&	\multicolumn{1}{c}{$\bullet$}	&		&		&		&		&		&	News	&	\multicolumn{1}{c}{$\bullet$}	&		\\
    
    \cite{ceran2015story}	&	2015	&		&		&		&	\multicolumn{1}{c}{$\bullet$}	&	\multicolumn{1}{c}{$\bullet$}	&		&	Other	&	\multicolumn{1}{c}{$\bullet$}	&		\\
    \cite{qamra2006mining} 	&	2006	&		&		&		&	\multicolumn{1}{c}{$\bullet$}	&	\multicolumn{1}{c}{$\bullet$}	&		&	Other	&	\multicolumn{1}{c}{$\bullet$}	&		\\
    \cite{bonchi2016identifying} 	&	2016	&		&		&		&	\multicolumn{1}{c}{$\bullet$}	&		&		&	Other	&	\multicolumn{1}{c}{$\bullet$}	&		\\
    \cite{Behrooz2015remember} 	&	2015	&		&		&		&		&		&		&	Other	&	\multicolumn{1}{c}{$\bullet$}	&		\\

    % \hline
    % \midrule
    \multicolumn{11}{ l }{\textit{Other articles}}\\
    % \hline	
    
    \cite{park2010detecting} 	&	2010	&		&		&		&	\multicolumn{1}{c}{$\bullet$}	&	\multicolumn{1}{c}{$\bullet$}	&	\multicolumn{1}{c}{$\bullet$}	&	Generic	&		&		\\
    \cite{kurashima2009discovering} 	&	2009	&		&		&		&	\multicolumn{1}{c}{$\bullet$}	&	\multicolumn{1}{c}{$\bullet$}	&	\multicolumn{1}{c}{$\bullet$}	&	Generic	&		&		\\
    \cite{inui2008experience} 	&	2008	&		&		&		&	\multicolumn{1}{c}{$\bullet$}	&	\multicolumn{1}{c}{$\bullet$}	&	\multicolumn{1}{c}{$\bullet$}	&	Generic	&		&		\\
    
    \cite{mazoyer2018realtime}	&	2018	&	\multicolumn{1}{c}{$\bullet$}	&		&		&	\multicolumn{1}{c}{$\bullet$}	&		&		&	News	&		&	\multicolumn{1}{c}{$\bullet$}	\\
    
    \cite{nanni2017building} 	&	2017	&	\multicolumn{1}{c}{$\bullet$}	&		&		&		&		&		&	News	&		&		\\
    \cite{kuzey2014fresh} 	&	2014	&	\multicolumn{1}{c}{$\bullet$}	&		&		&		&		&		&	News	&		&		\\
    
    \cite{lee2004multilingual}	&	2004	&	\multicolumn{1}{c}{$\bullet$}	&		&		&		&		&		&	News	&		&		\\
    \cite{brants2003system} 	&	2003	&	\multicolumn{1}{c}{$\bullet$}	&		&		&		&		&		&	News	&		&		\\
    
    \cite{khrouf2014mining}	&	2014	&		&		&		&	\multicolumn{1}{c}{$\bullet$}	&		&		&	Events	&		&		\\
    
    \cite{abe2011mining}	&	2011	&		&		&		&	\multicolumn{1}{c}{$\bullet$}	&	\multicolumn{1}{c}{$\bullet$}	&	\multicolumn{1}{c}{$\bullet$}	&	Other	&		&		\\
    \cite{kurashima2006mining} 	&	2006	&		&		&		&	\multicolumn{1}{c}{$\bullet$}	&	\multicolumn{1}{c}{$\bullet$}	&		&	Visiting	&		&		\\
    
    \cite{calix2017deep} 	&	2017	&		&		&		&	\multicolumn{1}{c}{$\bullet$}	&		&		&	Health	&		&	\multicolumn{1}{c}{$\bullet$}	\\
    \cite{jiang2017identifying} 	&	2017	&		&		&		&	\multicolumn{1}{c}{$\bullet$}	&		&	\multicolumn{1}{c}{$\bullet$}	&	Health	&		&	\multicolumn{1}{c}{$\bullet$}	\\
    \cite{hassan2016event} 	&	2016	&		&		&		&	\multicolumn{1}{c}{$\bullet$}	&		&		&	Product	&		&		\\
    \cite{krawczyk2016perceptual}	&	2016	&		&		&		&	\multicolumn{1}{c}{$\bullet$}	&		&		&	Hotel	&		&		\\
    \cite{masterov2015canary} 	&	2015	&		&		&		&	\multicolumn{1}{c}{$\bullet$}	&		&		&	Retail	&		&		\\
    \cite{liu2015context} 	&	2015	&		&		&		&	\multicolumn{1}{c}{$\bullet$}	&		&		&	Health	&		&		\\
    \cite{noferesti2015using}	&	2015	&		&		&		&	\multicolumn{1}{c}{$\bullet$}	&		&	\multicolumn{1}{c}{$\bullet$}	&	Patient	&		&		\\
    \cite{wilson2014finding} 	&	2014	&		&		&		&	\multicolumn{1}{c}{$\bullet$}	&		&	\multicolumn{1}{c}{$\bullet$}	&	Health	&		&	\multicolumn{1}{c}{$\bullet$}	\\
    \cite{hattori2012extracting} 	&	2012	&		&		&		&	\multicolumn{1}{c}{$\bullet$}	&		&	\multicolumn{1}{c}{$\bullet$}	&	Other	&		&		\\
    \cite{min2012identifying} 	&	2012	&		&		&		&	\multicolumn{1}{c}{$\bullet$}	&		&	\multicolumn{1}{c}{$\bullet$}	&	Product	&		&		\\
    \cite{garcia2010exploring} 	&	2010	&		&		&		&	\multicolumn{1}{c}{$\bullet$}	&		&	\multicolumn{1}{c}{$\bullet$}	&	Hotel	&		&		\\
    \cite{jijkoun2010mining} 	&	2010	&		&		&		&	\multicolumn{1}{c}{$\bullet$}	&		&	\multicolumn{1}{c}{$\bullet$}	&	Product	&		&		\\
    \cite{Vails-Vargas2017towards} 	&	2017	&		&		&		&		&		&		&	Other	&		&		\\
    \cite{goswami2016survey}	&	2016	&		&		&		&		&		&		&	Other	&		&		\\
    \cite{hattori2013tip}	&	2013	&		&		&		&		&		&		&	Other	&		&		\\

    \hline
    % \midrule
    \multicolumn{11}{ l }{Legend -- acronyms:}\\ 
    \multicolumn{11}{ l }{Nw: News \hspace{1.5em} SP: Software practice} \\
    \multicolumn{11}{ l }{PE: Professional experience \hspace{1em} Pl: Personal experience} \\
    \multicolumn{11}{ l }{Bl: Blog posts \hspace{0.5em} Em: Emotional experience \hspace{0.5em} St: Stories \hspace{0.5em} Tw: Twitter}\\
    % \hline
    % \botrule
    \end{tabular*}}{}
\end{table}

% \begin{table}
% \centering
%     % \small
%     % \caption{Classification of the retained articles}
%     % \label{table:classification-of-46-papers}
%     %\begin{tabular}{| p{0.6cm} | p{0.7cm} | p{0.7cm} | p{0.7cm} | p{0.7cm} | p{0.7cm} | p{0.7cm} | p{0.7cm} | p{0.7cm} | p{1.8cm} | p{0.7cm} | p{0.7cm}|}

%     \processtable{\textcolor{blue}{Frequency of publication of papers}\label{table:frequency-of-publication}}
%     {\begin{tabular*}{20pc}{@{\extracolsep{\fill}} l c @{}}
%     \toprule
%     Year &	Frequency \\
%     \midrule
%     2018 &	2 \\
%     2017 &	7 \\
%     2016 &	4 \\
%     2015 &	6 \\
%     2014 &	4 \\
%     2013 &	1 \\
%     2012 &	4 \\
%     2011 &	2 \\
%     2010 &	5 \\
%     2009 &	1 \\
%     2008 &	1 \\
%     2007 &	0 \\
%     2006 &	3 \\
%     2005 &	0 \\
%     2004 &	1 \\
%     2003 &	1 \\
%     Total &	42 \\
%     \botrule
%     \end{tabular*}}{}
% \end{table}

\begin{figure}[ht]
  \centering
    \includegraphics[scale=0.6]{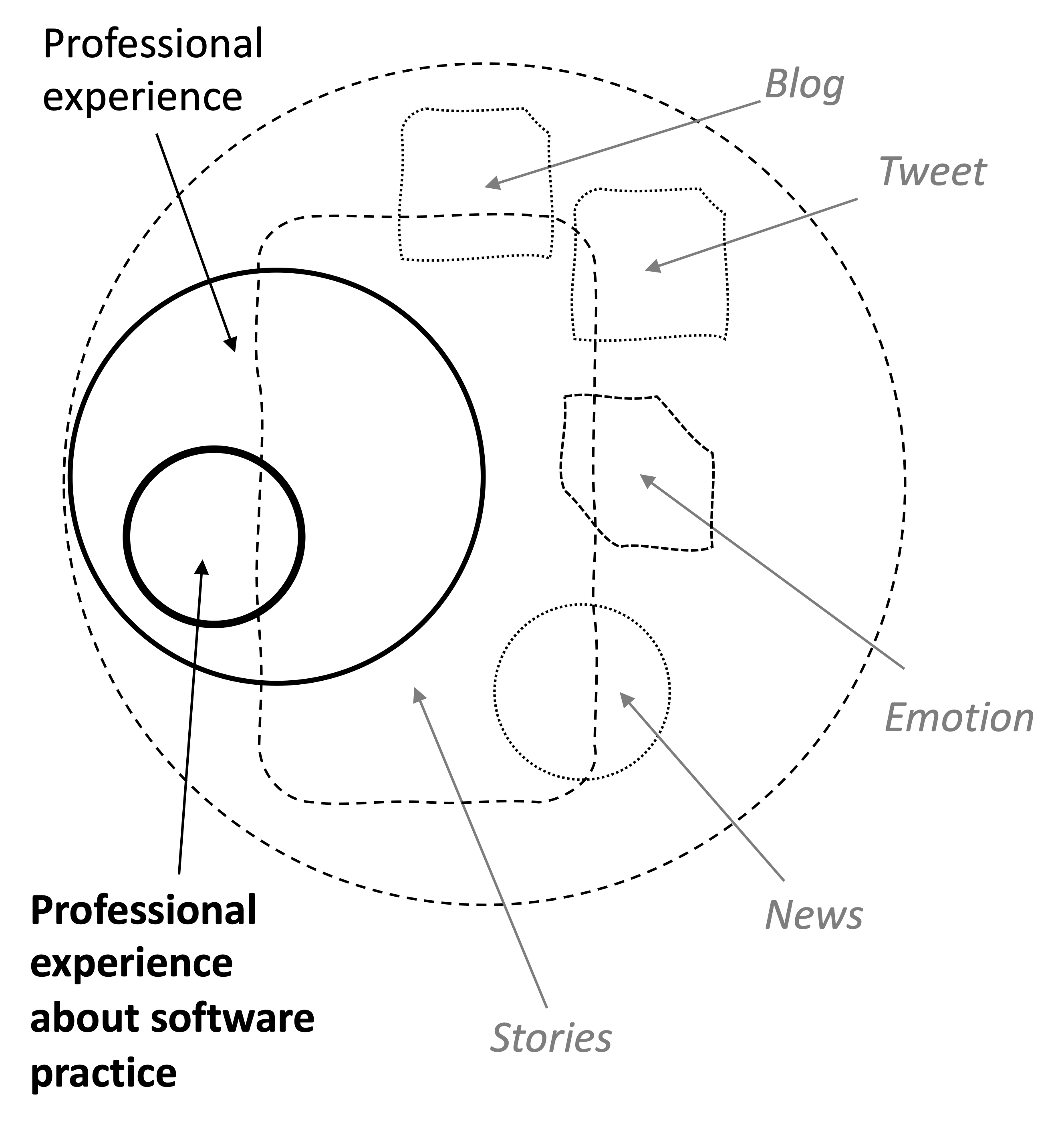}
  \caption{An illustration of the `experience space'}
  \label{figure:experience-space}
\end{figure}

\textcolor{black}{Figure \ref{figure:experience-space} visualises the \textit{multi--dimensional} `experience space' characterised by Table \ref{table:classification-of-46-papers}. The figure illustrates the challenges of conducting retrieval and mining that \textit{include} online articles about the professional experience of software practice whilst  \textit{excluding} articles about a number of other closely--related issues.}

\textcolor{black}{Our overall conclusion is that whilst there has been more than 15 years' interest in mining experience, it appears that there has been limited work published in this area. More specifically, there appears to be a very limited number of studies -- and arguably no previous study -- of the retrieval and mining of self--reports of professional experience of software practice. This suggests a nascent field of research that has yet to establish itself and that remains indistinguished from already--established fields such as opinion mining \cite{hemmatian2019survey}, event detection \cite{chen2019bibliometric} and argumentation mining \cite{lippi2016argumentation}.}

In the following section we consider some challenges likely to arise during the retrieval and mining of self--reports of professional experience of software practice.
\section{Analysis of challenges}
\label{section:analysis}

In this section we discuss several challenges and issues that arise with the retrieval and mining of online articles that contain reports of experience. \textcolor{black}{The eight papers considered in this section are summarised in Table \ref{table:first-classification-of-nine-papers}.}

\subsection{Characteristics of experience--reporting articles}
\label{subsection:characteristics}

Jijkoun \textit{et al}. \cite{jijkoun2010mining} found that online articles reporting experience are different to those that do not report experience: experience--articles have twice as many sentences, twice as many words (these two statistics are clearly related), twice as many references to the pronoun \textit{I}, and twice as many non--modal verbs. There is, therefore, an indication that experience--reporting articles are different to non--experience articles, however we are unaware of any other study that has reported such characteristics and, as indicated by our visualisation in Figure \ref{figure:experience-space}, we further distinguish experience--reporting articles from professional--experiencing--reporting articles.

\subsection{The prevalence of experience in online articles}
\label{subsection:prevalence}

There is very limited information reported in the articles on the prevalence of experience in online articles.  Swanson \textit{et al}. \cite{swanson2014identifying} cite their previous work \cite{gordon2009identifying} that found about 5\% of blog posts are about personal stories. Ceran \textit{et al}. \cite{ceran2012hybrid} found 3,301 story--paragraphs in a dataset of 16,930 (19\%; though these are paragraphs rather than documents). On the basis of this limited information, it appears that there would likely be a low prevalence of reporting professional experience in online articles. This degree of prevalence increases the challenges of generating appropriate datasets, of classifier training and evaluation, and then of effective application of trained classifiers e.g. to automate aspects of grey literature reviews (GLRs) and multivocal literature reviews (MLRs).

\subsection{Definitions and models of experience, and theories}
\label{subsection:theory}

Table \ref{table:first-classification-of-nine-papers} suggests that a limited number of papers base their work on prior theory, though most papers have some kind of model of experience to inform the selection of features. The most frequent model of professional experience appears to be in terms of some kind of story, though there are several different types of story being used by research e.g. news stories, folk stories. Swanson \textit{et al}. \cite{swanson2014identifying} asserts that storytelling is considered a fundamental aspect of human social behaviour. Swanson \textit{et al}. \cite{swanson2014identifying} also considered that the narrative structures of personal narratives posted in blog posts are likely to be more similar to oral narratives than to classical stories, and consequently chose Labov and Waletzky's \cite{labov1967narrative} theory of oral narrative to distinguish causal relationships.  By contrast, Goncalves \textit{et al}. \cite{de2010case} refer to story mining and story telling, but use a metamodel of scenarios \cite{achour1999guiding} to model experience rather than a model of stories. Table \ref{table:elements-in-story-based-models-of-experience} contrasts the conceptual elements of the different models used in our sample, and Table \ref{table:challenges-modelling-experience} summarises some of the challenges of `mapping' the conceptual elements to linguistic elements of a text.  Aside from datasets, the keywords to use for retrieving online articles, and then the features to use (e.g., in machine classifiers) to mine or classify the articles, there is also the conceptual challenge of modelling the reporting of experience in online articles.

\begin{table*}[ht]
\centering
\small
% \caption{Summary classification of the subset of papers}
% \label{table:first-classification-of-nine-papers}
\caption{Summary classification of the subset of papers\label{table:first-classification-of-nine-papers}}
{\begin{tabular*}{38pc}{p{1.5cm} c p{0.6cm} p{4cm} c p{1.5cm} c c}
% \toprule
% \begin{tabular}{|p{1.5cm} | p{0.6cm} | p{0.6cm} | p{4cm} | p{1.8cm} | p{1.5cm} | p{0.7cm} | p{1.2cm} |}
% \hline
\multirow{2}{1.5cm}{\textbf{First author}} & \multirow{2}{0.4cm}{\textbf{Ref}} & \multirow{2}{0.8cm}{\textbf{Date}} & & \textbf{Model of} & \textbf{Unit of} & \multirow{2}{0.7cm}{\textbf{Story}} & \multirow{2}{1.2cm}{\textbf{Pipeline}}\\
& & & \textbf{Prior theory} & \textbf{exp'nce} & \textbf{discourse} &  & \\
%\textbf{ID} & \textbf{First author} & \textbf{Ref} & \textbf{Date} & \textbf{theory} & \textbf{experience} & \textbf{discourse} & \textbf{Story} & \textbf{Pipeline}\\
\hline
\hline
% \midrule
Hassan & \cite{hassan2016event} & 2016 & No & No & Sentence & No & Yes\\
% \hline
Swanson & \cite{swanson2014identifying} & 2014 & Yes: Labov \& Waletzky \cite{labov1967narrative} & Yes &  Clause& Yes & No \\
% \hline
Gon{\c{c}}alves & \cite{de2010case} & 2010 & Implied (see \cite{de2011collaborative}): CREWS \cite{achour1999guiding} & Yes & Sentence & Yes & Yes\\
% \hline
Park & \cite{park2010detecting} & 2010 & Yes: Vendler \cite{vendler1967linguistics} & Yes & Sentence & No & No\\
% \hline
Kurashima & \cite{kurashima2009discovering} & 2009 & No & Yes & Document & No & No\\
% \hline
Inui & \cite{inui2008experience} & 2008 & No & Yes & Sentence & No & No\\
% \hline
Kurashima & \cite{kurashima2006mining} & 2006 & No & No & Sentence & No & Yes\\
% \hline
% \hline
% \midrule
Ceran & \cite{ceran2012hybrid}  & 2012 & No & Yes & Sentence & Yes & Yes\\
% \botrule
% \hline
% \textcolor{blue}{\textit{Gon{\c{c}}alves}} & \textit{\cite{de2011collaborative}} & \textit{2011} & \textit{Yes \cite{achour1999guiding}} & \textit{Yes} & --- & \textit{Yes} & \textit{No} \\
% \hline
% \hline
% \multicolumn{8}{|l|}{An en dash denotes the source paper does not report the information}\\
% \hline
\end{tabular*}}{}
\end{table*}

\begin{table*}[ht]
\centering
\small
% \caption{Summary of the datasets and their annotation}
% \label{table:classification-of-datasets-in-the-nine-papers}
\caption{Summary of the datasets and their annotation\label{table:classification-of-datasets-in-the-nine-papers}}
{\begin{tabular*}{38pc}{p{2.6cm}  p{8.8cm}  p{1.5cm}   p{0.3cm}  p{0.4cm}}
% \toprule
% \begin{tabular}{| p{1.6cm} | p{0.6cm} | p{10cm} | p{1.4cm} |  p{0.4cm} | p{0.4cm} |}
% \hline
% \textbf{First} &\multicolumn{2}{ c }{\textbf{}} & \multicolumn{2}{ c } {\textbf{Annotation}}\\
%\cline{3-6}
\textbf{Author} & \textbf{Studied size, overall size, and source/s} & \textbf{Lang.} & \textbf{\#} &  \textbf{$\kappa$}\\
\hline
\hline
% \midrule
 Hassan  \cite{hassan2016event} & 383 true reviews (268 training, 115 test) from 400 true reviews \cite{ott2011finding} & English & 1 & --\\
% \hline
Swanson \cite{swanson2014identifying} & 50 personal stories drawn from 5000 posts \cite{gordon2009identifying} taken from 44M articles \cite{burton2009icwsm} & English & 3 & 0.58\\
% \hline
% \textcolor{blue}{\textit{Gon{\c{c}}alves}} \cite{de2011collaborative} & ---  & --- & --- & ---\\
% \hline
Gon{\c{c}}alves \cite{de2010case} & One case selected from stories generated for the study, relating to one business process & Portug'se & -- & --\\
% \hline
Park \cite{park2010detecting} & 588 sentences from 6000 blog posts on Wordpress & English & 3 & -- \\
% \hline
Kurashima \cite{kurashima2009discovering} & 29M blog posts from 48M blog posts collected using the Blogranger 2.0 API & -- & 3 & -- \\
% \hline
Inui \cite{inui2008experience} &  50M posts, related to pre--selected topics, from 150M weblog posts [No stated source] & Japanese & 2 & 0.68\\
% \hline
Kurashima \cite{kurashima2006mining} & 62,396 articles from Two Japanese blog hosting sites & Japanese & -- & --\\
% \hline
% \hline
% \midrule
Ceran  \cite{ceran2012hybrid} & 16,930 paragraphs from 1,256 documents (13629 non--story, 3301 story)[No stated source] & English & -- & 0.83\\
% \hline
% \hline
% \botrule
% \multicolumn{6}{l}{}\\
\hline
\multicolumn{5}{l}{An em dash denotes the source paper does not report the information}
\end{tabular*}}
\end{table*}

\begin{table*}[!ht]
\small
\caption{Elements of conceptual models of experience}
\label{table:elements-in-story-based-models-of-experience}
{\begin{tabular*}{14cm}{p{2cm} p{12cm} }
% \toprule
% \midrule
\multicolumn{2}{ l }{\textit{Elements of story based models of experience}}\\
\hline
\hline
Swanson \cite{swanson2014identifying} & \textit{Action}: clauses describing causal or event relationships. A story must have an \textit{actor} or actors. \textit{Orientation}: clauses describing properties about settings (e.g. time and location) and actors. \textit{Evaluation}: clauses describing emotional reactions, and the reason for telling the story. \textit{Abstract}: an initial clause summarising the entire sequence of events. \textit{Coda}: a final clause ending the story and providing a 'moral' to the story.\\
Gon{\c{c}}alves \cite{de2010case} & Uses a subset of the CREWS scenario meta--model \cite{achour1999guiding}, focusing on Activities (\textit{Actions} and \textit{Flows of actions} in CREWS) and Actors (\textit{Objects} in CREWS).\\
Ceran \cite{ceran2012hybrid} & A story must have an \textit{actor} or actors. The actor/s must be performing \textit{actions}. The actions must result in a \textit{resolution}.\\
\multicolumn{2}{ l }{}\\
\multicolumn{2}{ l }{\textit{Elements of non--story based models of experience}}\\
\hline
\hline
Park \cite{park2010detecting} & The authors state that experiences can be recorded in terms of attributes, such as: \textit{location}, \textit{time} and \textit{activity} and their \textit{relations}. They focus on two of Vendler's \cite{vendler1967linguistics} four basic classes of verbs: \textit{states} and \textit{activities} (excluding \textit{achievements} and \textit{accomplishments}).\\
Inui \cite{inui2008experience} & Each experience is represented as a piece of structured information comprising: A person's experience is expressed by five attributes: \textit{Topic}: what the experience is about. \textit{Experiencer}: the person experiencing the event. \textit{Event expression}: The event that is experienced.  \textit{Event type}: The semantic type of the event (not defined in the paper). \textit{Factuality}: Whether the event indeed took place i.e the temporal and modal status of the event \textit{Source pointer}: a reference to the source text \\
Kurashima \cite{kurashima2009discovering} & A person's experience is expressed by five attributes: \textit{Time}: when a person (actor) acted. \textit{Location}: Where the person acted. \textit{Activity}: An action and its object. \textit{Opinion}: an evaluation about the object. \textit{Emotion}: The feeling of the person.\\
% \botrule
\end{tabular*}}{}
\end{table*}

% \begin{table}[!b]
% \processtable{Example table\label{tab1}}
% {\begin{tabular*}{20pc}{@{\extracolsep{\fill}}lll@{}}\toprule
% Column  &Column  & Column heading \\
% heading  &heading two &  three \\
% \midrule
% Row 1a  &Row 1b  &Row 1c \\
% Row 2a  &Row 2b  &Row 2c \\
% Row 3a  &Row 3b  & Row 3c \\
% Row 4a  &Row 4b  &Row 4c \\
% Row 5a  &Row 5b  &Row 5c \\
% Row 6a  & Row 6b  & Row 6c \\
% \botrule
% \end{tabular*}}{}
% \end{table}

\begin{table*}[ht]
\centering
\small
% \caption{Challenges of mapping conceptual elements to linguistic elements (for story--based experience)}
% \label{table:challenges-modelling-experience}
% \begin{tabular}{| p{2cm} | p{0.1cm} p{14cm} |}
\caption{Challenges of mapping conceptual elements to linguistic elements (for story--based experience)\label{table:challenges-modelling-experience}}
% {\begin{tabular*}{13.1cm}{p{2cm} p{0.1cm} p{11cm}}
{\begin{tabular*}{13.1cm}{p{0.1cm} p{13cm}}
% \toprule
% \hline
% \textbf{First author} & \multicolumn{2}{ l }{\textbf{Challenge/s}}\\
% \hline
% \hline
% \midrule
\multicolumn{2}{ l }{}\\
\hline
\multicolumn{2}{ l }{Swanson \cite{swanson2014identifying} \textit{et al}.  identify six primary sources of annotator disagreement:}\\
1 & Clauses of more than one category are common with types of \textit{action} and \textit{evaluation} and e.g. containing elements of \textit{orientation}, \textit{action} and \textit{evaluation}.\\
2 & There are \textit{actions} implied but not explicitly stated in the text.\\
3 & Stative descriptions of the world, as a result of an \textit{action}, are not intuitively \textit{orientation}.\\
4 & Stative descriptions of the world that are localised to a specific place in the narrative does not easily align with \textit{orientation}\\
5 & Disambiguating the functional purpose of clauses that describe \textit{actions}, but may be intended to be \textit{orientation}\\
6 & Disambiguating the functional purpose of subjective language in the description of an event or state e.g. the \textit{gigantic} tree, the \textit{radiant} blue of the sky.\\
% \hline
% \midrule
\multicolumn{2}{ l }{}\\
\hline
\multicolumn{2}{ l  }{Ceran \cite{ceran2012hybrid} \textit{et al}. make the following inferences:}\\
1 & Stories will have a lower proportion of stative verbs than non--stories, because stories describe actions.\\
2 & Stories will include more named entities, especially person names, than non--stories.\\
3 & Stories will use more personal pronouns than non--stories. \\
4 & Stories may repeat similar nouns. \\
5 & Paragraphs with stories in them will have different sentence lengths than paragraphs without stories [it is not clear whether sentences of story--paragraphs would be longer or shorter.]\\
% \hline
% \botrule
\end{tabular*}}{}
\end{table*}

\subsection{The spareness of data}
\label{subsection:data-spareness}
% Theories and models of personal experience, and the implications arising from those theories and models (such as the narrative structures of stories, the type of story, the unit of discourse) influence the nature of the dataset/s to use and the choice of features.

%Goncalves \textit{et al}. \cite{de2010case}: The study is different to others however in that the participants \textit{generate stories} for the system. All other studies focus on working with previously generated data that has been generated for other reasons.

Kurashima \textit{et al}. \cite{kurashima2009discovering} acquired a large dataset of 48M blog posts, but found that 38\% of the dataset was not suitable because it did not contain a sufficient co--occurrence of words they needed for their association rules. The authors observed a spareness in their data e.g. the average frequency of their \textit{activity} words per blog post was 4.8; for \textit{location} words the average frequency was 0.85, and for emotion words the average frequency was 0.26. 

\subsection{Unit/s of discourse analysis}
\label{subsection:units-of-discourse}

Table \ref{table:first-classification-of-nine-papers} suggests that the majority of studies use the sentence as the basic unit of discourse. This is consistent with the findings of Lippi and Torroni \cite{lippi2016argumentation} in their review of argumentation mining. Reporting of professional experience will likely fall across sentences and paragraphs (see for example the analysis in Rainer \cite{rainer2017using}) so there will likely be the challenge of appropriate text segmentation and aggregation.

\subsection{Annotating, and annotated datasets}
\label{subsection:annotation}

Table \ref{table:classification-of-datasets-in-the-nine-papers} shows that only three of our studies report measures of annotation agreement.  Swanson \textit{et al}. state, ``\dots\space the annotation task is highly subjective, requiring interpreting the narrative and the author's intention, which prevents us from obtaining high levels of inter--rater agreement.'' (\cite{swanson2014identifying}; p. 175). They also observe earlier in their paper that a previous study \cite{rahimtoroghi2013evaluation} found both a high level of annotator agreement and an extremely high machine learning accuracy for Aesop's Fables. Swanson \textit{et al}. \cite{swanson2014identifying} infer that the `classical', written--down stories are therefore (much) easier to work with than blog--posts. This observation is consistent with the progress made by argumentation mining i.e. to first begin by mining texts that have a consistent structure and content e.g. legal texts. Swanson \textit{et al}. \cite{swanson2014identifying} also recogise that a more sophisticated analytical framework, and annotation scheme, such as that of DramaBank has advantages, but developing an annotated corpus using such an analytical framework for blog posts would be (highly) resource intensive.

Kurashima \textit{et al}. \cite{kurashima2009discovering} focused on two types of association rules. For the first type, they identified 1659 candidate rules, for which their annotators confirmed 129 correct rules (8\%); and for the second type, they identified 1330 candidates rules, for which their annotators confirmed 55 correct rules (4\%). Again, this suggests a considerable challenge with annotating the data and then mapping the annotated data to the classified data. 

%\subsection{Mapping text to model}

%Identify the contiguous `content' of interest is a challenge e.g. the story, or the experience. This is partly a problem of `boundary detection' or 'content' segmentation. To simplify the challenge, research tends to focus on a relatively low--level unit of discourse e.g. the sentence or the clause, and then look for some unit--level content.

\subsection{Evaluating the performance of algorithms}
\label{subsection:evaluation}

% Similar to Lippi and Torroni's \cite{lippi2016argumentation,lippi2016margot}) pipeline for argumentation mining, we observe a process of:
% \begin{enumerate}
%     \item Acquire (or generate) an appropriate dataset(s);
%     \item Extract units of discourse (e.g. sentences) from a text;
%     \item Extract properties of those units e.g. parts of speech;
%     \item Infer elements of experience from the (properties of the) units of discourse e.g. infer elements of a story from parts of speech;
%     \item Infer the presence of experience from the aggregated (and inferred) presence of elements; and
%     \item Infer the presence of an appropriate experience.
% \end{enumerate}

% Evaluation can occur at each step in this process, and at different levels of granularity, e.g. evaluating an algorithm for extracting grammatical structures from sentences in comparison to evaluating an overall tool, or evaluating the effectiveness of segmenting experience sentences in contrast to predicting (aggregating) an overall experience from the experience sentences. Also, of course, one can distinguish between the effectiveness of the classifying system in contrast to the practical value of that system for use by researchers or practitioners (cf. statistical significance vs practical significance).

Different articles report different approaches to retrieving and mining their data, and this makes it difficult to compare the performance of different approaches. We contrast three examples here.

Swanson \textit{et al}. (\cite{swanson2014identifying}  investigate performance both in terms of (linguistic) clauses and also stories. They observe that, ``\dots\space some clauses are ambiguous and difficult [for the annotators] to label, but also that some \textit{stories} are more difficult to classify.'' (\cite{swanson2014identifying}, p. 177); and, ``Performance is different for results by story rather than over all clauses. This indicates that some stories are more difficult to classify than others and that ambiguous clauses are not uniformly distributed but are likely to be correlated with particular authors or writing styles.'' (\cite{swanson2014identifying}, p. 178). Swanson \textit{et al}. \cite{swanson2014identifying} is the only article to present a confusion matrix; also they discussed sources of error (which we do not consider here, due to space constraints). 

By contrast, Goncalves \textit{et al}. \cite{de2010case} evaluate the overall tool--process, with one case example, rather than the method by which linguistic features are mapped to process attributes. As another contrast, Kurashima \textit{et al}. \cite{kurashima2009discovering} sought to identify the more `interesting' association rules, and were also seeking an appropriate measure for 'interestingess'.

\subsection{Replications}
\label{subsection:replication}
In our dataset of articles, we found no replication, independent or otherwise, of previous work and no attempts to independently validate the findings, or assumptions, of others' work.
\section{Discussion}
\label{section:discussion}

\subsection{A review of the paper's objectives}

Recall from section \ref{section:introduction} that the two objectives for the paper were: 1) to classify the articles in our dataset so as to better understand prior work in this area; and 2) to identify challenges and issues with the retrieval and mining of professional experience of software practice.

For our first objective, we identified and classified 42 articles. In so doing, we showed that there were very few previous studies investigating the retrieval and mining of professional experience of software practice. 

% We found as well that:
% \begin{enumerate}
%     \item The first articles published in this area appear to be Kurashima \textit{et al}. \cite{kurashima2006mining} in 2006, followed by Inui \textit{et al}. \cite{inui2008experience} in 2008.
%     \item There appears to be little work conducted in this area over the years.
%     \item Story is a common approach to express experience.
%     \item There are many more articles that report on non--professional (personal) experience than professional experience.
%     \item Blogs and tweets were the main social media explored in previous studies.
%     \item Some articles focus on qualities of experience, such as emotion.
%     \item It is, or will be, challenging to retrieve and mine online articles that \textit{include} only online articles about the professional experience of software practice whilst \textit{excluding} online articles about a number of other closely--related issues.
% \end{enumerate}

For our second objective, \textcolor{black}{we  found that it will be very challenging to retrieve and mine \textit{only} those online articles that are about the professional experience of software practice whilst \textit{excluding} online articles about a number of other closely--related issues. Challenges include}:
\begin{enumerate}
    % \item There appears to be characteristics of articles that distinguish articles reporting experience (not necessarily professional experience) from those articles not reporting experience, however there is little prior work in this area.
    \item There is a low prevalence of experience (not necessarily professional experience) reported in online articles, the implication being that a very large dataset would be needed to, for example, train and test automated tools.
    \item There is no consistently used definitions, models or theories of experience. A consistent definition etc. would help with, as examples, replication and the accumulation of evidence. Conversely, contrasting definitions etc. are important for exploration. Overall, the lack of consistent use of definitions etc. suggests the field is still maturing. In our sample, story is the most common approach to articulate experience however story is defined in different ways by different researchers \textcolor{black}{(see Table \ref{table:challenges-modelling-experience} for examples)}.
    \item Datasets are sparse of features, i.e., had very low frequencies of the presence of features.
    \item There is variation in the units of discourse analysis used, with the sentence being the most common unit. It is also not clear how a text would be segmented for experience.
    \item Previous work tends not to report annotator agreement and, where annotator agreement is reported, the authors recognise considerable difficulties achieving agreement.
    \item There is considerable variation in the use of approaches and algorithms, and difficulty comparing their performance.
    \item There has been no replication of previous research.
\end{enumerate}

Overall, our findings are consistent with those of O’Mara-Eves \textit{ et al}. \cite{OMara-Eves2015} and Cabrio and Villata \cite{cabrio2018five} in their respective research domains.

\subsection{\textcolor{black}{Guiding questions for primary studies}}

\textcolor{black}{Given the lack of prior research in this area, and the challenges of even identifying and then comparing studies from prior research, we propose a simple set of guiding questions for the design, conduct and reporting of the retrieval and mining of professional experience. Space prevents a detailed discussion of these questions. We present the questions in Table \ref{table:framework-for-definitions}, with brief cross--reference to relevant subsections earlier in paper.}

% As noted in section \ref{section:introduction} we began with working definitions that would be reviewed following the classification of articles, and the identification of challenges and issues. Given the variation in the articles we have classified and analysed, we think it more valuable to suggest a simple set of guidelines for future primary studies in this area, and relate our working definitions to those guidelines. Future research can use the guidelines to inform the design of their studies. The guidelines are presented in Table \ref{table:framework-for-definitions}. Space prevents a more detailed discussion of these guidelines.

\begin{table}[ht]
\centering
\small
\caption{Guiding questions for primary studies investigating the mining of professional experience\label{table:framework-for-definitions}}
\scalebox{0.85}{\begin{tabular*}{14cm}{p{0.3cm} p{13cm}}
% \toprule
% \begin{table}[ht]
% % \centering
% \small
% \caption{Preliminary guidelines for primary studies}
% \label{table:framework-for-definitions}
% \begin{tabular}{ p{0.3cm} p{7cm}}
% \hline
% \toprule
\textbf{\#} & \textbf{Question \textcolor{black}{[with cross--reference to section \S, Table or Figure]}}\\
% \midrule
\hline
\hline
\multicolumn{2}{l}{\textit{Theory}}\\
% \hline
T1 & How is the study informed by an appropriate theory of experience? A theory would support generalisation and/or aggregation of independently conducted studies. \textcolor{black}{[cf. \S \ref{subsection:theory}]} \\
T2 & \textcolor{black}{How is the study positioned relative to, and distinct from, fields such as opinion mining, event detection, argument mining etc.? [cf. \S \ref{section-related-work}]}\\
% \multicolumn{2}{l}{}\\
\hline
\multicolumn{2}{l}{\textit{Definitions}}\\
% \hline
D1 & What is your definition of professional experience of software practice? \textcolor{black}{[cf. \S \ref{subsection:theory}]}\\
D2 & Is your definition derived from theory? \textcolor{black}{[cf. T1 and \S \ref{subsection:theory}]}\\
D3 & Do you have appropriate sub--definitions of experience, e.g., different domains of software practice, different domains of experience? \textcolor{black}{[cf. \S \ref{subsection:theory}, Figure \ref{figure:experience-space}]}\\
\hline
% \multicolumn{2}{l}{}\\
\multicolumn{2}{l}{\textit{Representation of experience}}\\
% \hline
R1 & What representation, or representations, of experience will you seek to retrieve and mine? \textcolor{black}{[cf. \S \ref{subsection:theory}]} Story is a common representation, but even story is defined and measured in different ways.\\
\hline
% \multicolumn{2}{l}{}\\
\multicolumn{2}{l}{\textit{Natural language}}\\
% \hline
L1 & What natural language or languages will you retrieve and mine,e.g., English, Japanese, Portuguese? \textcolor{black}{[cf. Table \ref{table:classification-of-datasets-in-the-nine-papers}]. Research appears to have focused more on English and Japanese.}\\
\hline
% \multicolumn{2}{l}{}\\
\multicolumn{2}{l}{\textit{Social media}}\\
% \hline
S1 & From what social media will you retrieve information? \textcolor{black}{[cf. \S \ref{subsection:prevalence}]}\\
\hline
% \multicolumn{2}{l}{}\\
\multicolumn{2}{l}{\textit{Units of discourse}}\\
% \hline
U1 & What is your basic unit/s of discourse, e.g., sentence, clause, paragraph? \textcolor{black}{[cf. \S \ref{subsection:units-of-discourse}]}\\
\hline
% \multicolumn{2}{l}{}\\
\multicolumn{2}{l}{\textit{Measures and features}}\\
% \hline
M1 & How will you map your definition of experience, and representation of experience (e.g., story) to measures and features in the dataset? \textcolor{black}{[cf. \S \ref{subsection:characteristics}]}\\
\hline
% \multicolumn{2}{l}{}\\
\multicolumn{2}{l}{\textcolor{black}{Annotation}}\\
% \hline
A1 & \textcolor{black}{How will you design, conduct and report the annotation process? [cf. \S \ref{subsection:annotation}]} \\
A2 & \textcolor{black}{How will you handle disagreements during annotation?}\\
% \multicolumn{2}{l}{}\\
\hline
\multicolumn{2}{l}{Technologies}\\

X1 & What technologies (e.g., natural language tools) will you use and how will you report those in a way that allows future replication, evaluation and comparison? \textcolor{black}{[cf. \S \ref{subsection:evaluation}, \ref{subsection:replication}]}\\
\hline
% \multicolumn{2}{l}{}\\
\multicolumn{2}{l}{\textit{Open science}}\\
% \hline
O1 & How will you make your study artefacts (e.g., source data) available for reuse by others, including replication? \textcolor{black}{[cf. \S \ref{subsection:replication}]}\\
% \hline
% \botrule
\end{tabular*}}{}
\end{table}

\subsection{\textcolor{black}{Reporting negative results in secondary studies}}

\textcolor{black}{One recurring item of feedback on earlier versions of this paper may be summarised with the following question:  What is the contribution of a negative result in a secondary study?}

\textcolor{black}{Primary and secondary studies both contribute knowledge to a field or fields of research.  A secondary study is often also an empirical study in its own right, albeit that the \textit{unit of data} for the secondary study (e.g., information from primary studies) is different to that for the primary study (e.g., information direct from practitioners or observations of the phenomena of interest itself). But both types of study follow methodology to analyse data and to interpret the results of that analysis. And both types of study may report false--negative and false--positive results.}

\textcolor{black}{Because primary studies reporting positive results are much more frequently published, secondary studies are often unavoidably `loaded' toward analysing primary studies that report positive results. We suggest this recurring situation leads to a normalisation of the expected results of a secondary study and to a subtle assumption as to what secondary studies \textit{should} report, i.e., that a defining characteristic of a secondary study is that it reports `positive' results about a field of research (and, by implication, that a field of research must have positive results for reporting in a secondary study). }

\textcolor{black}{The first and second authors of the current paper were interested in experience mining because of how it relates to research they were conducting on the credibility of grey literature. Our initial motivation, therefore, was not to advance the field of experience mining (either through conducting a primary study or a secondary study) but rather to use existing knowledge from the field. With several related fields of research already relatively well--developed (e.g., opinion mining, event detection and argumentation mining) we hoped to find an existing secondary study on experience mining, and we assumed we would find at least some primary studies on experience mining. (In other  words, \textit{our} expectations were for positive results.) When we didn't find a secondary study, we decided to conduct one. In conducting our secondary study we found a lack of primary studies.}

\textcolor{black}{One counter--argument is that we might have quickly detected the lack of primary studies early, or earlier, in the conduct of our secondary study, e.g., some `quick' Google Scholar searches would have shown there was little prior work. As we report in section \ref{section:method}, our initial search returned 300 results and, as we progressed in the study, our preliminary selection and rejection process (see Table \ref{table:search-queries}) continued to suggest we would find at least some relevant primary studies. A risk with making a decision based on the quick search for results is that we terminate the study prematurely, i.e.. we make a decision based on false--negative results. One of the principles of systematic reviews is that the reviewer performs a thorough and balanced search for relevant literature. One potential outcome of a balanced secondary study is to report negative results.}

\textcolor{black}{As noted in the introduction, an alternative strategy might be (have been) to relax the search and selection criteria so as to include more studies for review. For example, we might have expanded our secondary study to look at mining in general, drawing on argument mining, opinion mining etc. Whilst relaxing the search strategy would have widened the number of primary studies, the relaxed strategy would also have diverted us from the focus of our research, i.e., to understand how \textit{experience mining} might help us with assessing the credibility of grey literature. Also, relaxing our search strategy would change the objectives of the review and, by so doing, would fail to report a true negative and potentially contribute to publication bias.}

\subsection{\textcolor{black}{Threats to validity}}

There are of course threats to the validity of our review. The conduct of the searches and the initial selection of articles was performed by one author, the second author. There may be bias in that search and selection process. We might have applied a more extensive set of search terms and search strategies (including snowballing) to ensure we found all relevant literature to review. Restricting our review to the first 50 results may have limited the articles we found, though we found that a very small percentage of articles were relevant in the first 50 results and we did conduct a series of searches using a number of different academic search engines. There are also threats relating to the application of the inclusion and exclusion criteria, as well as other aspects of the selection process. The classification framework was prepared by one author, the first author. This framework was applied independently by the other authors, with all articles being classified independently by at least two authors, and most articles being classified independently by three.

\subsection{\textcolor{black}{Further research}}

There are several directions in which to extend this research. First, an unresolved and fundamental challenge for mining professional experience for GLRs and MLRs in software practice is establishing the appropriate features to use when classifying experience (cf. \cite{domingos2012few}). We observe in our review a lack of consensus on appropriate features and, of course, features may be domain specific. Second, there is also the challenge of generating appropriate datasets. Garousi {et al}. \cite{garousi2016and} and Soldani {et al}.\cite{soldani2018pains} provide two datasets as starting points, however both are very small (each approximately 50 articles) for training classifiers. In addition to the size of the dataset, there is the related issue of the spareness of data. A third challenge is the effective reporting of the full details of a study, so as to support reproducibility and replication. A fourth direction is to apply lessons learned and technology from the retrieval and  mining of other domains to the retrieval and mining of software practice. Finally, there is the need to evaluate the effective contribution that experience mining could make in due course to secondary studies, such as GLR (e.g. \cite{soldani2018pains}) and MLRs (e.g. \cite{garousi2018guidelines}), and to primary studies (e.g. \cite{parnin2011measuring}). Recall that O’Mara-Eves \textit{ et al}. \cite{OMara-Eves2015} found \textit{text mining} to be ready to use for \textit{screening papers} in their domain.

\subsection{\textcolor{black}{Conclusions}}

The overall aim of this paper was to review articles that investigate how to retrieve and mine from online articles the self--reporting of professional experience of software practice.  We set two objectives: 1) to classify the articles in our dataset; and 2) to identify challenges. 

Our review identified only \textit{one} directly relevant article and this single article concerns the software professional's emotional experiences rather than the software professional's reporting of behaviour and events occurring during software practice. It appears that no prior directly--relevant research has been conducted in this area: experience mining is a nascent field of research. We complement our near--null result with a classification of the studies, and with the identification of several challenges. We also propose a simple set of guiding questions to support future researchers. We discuss the value of negative results in secondary studies.

\section*{Acknowledgments}\label{sec11}

We thank Dionysis Athanasopoulos for his independent classification of a subset of the articles presented in Table \ref{table:classification-of-46-papers}.

\printbibliography

@inproceedings{rainer2003persuading,
  title={Persuading developers to ``buy into'' software process improvement: a local opinion and empirical evidence},
  author={Rainer, Austen and Hall, Tracy and Baddoo, Nathan},
  booktitle={Proceedings of the 2003 International Symposium on Empirical Software Engineering (ISESE 2003)},
  pages={326--335},
  year={2003},
  organization={IEEE}
}

@inproceedings{devanbu2016belief,
  title={Belief \& evidence in empirical software engineering},
  author={Devanbu, Prem and Zimmermann, Thomas and Bird, Christian},
  booktitle={Proceedings of the 38th International Conference on Software Engineering},
  pages={108--119},
  year={2016},
  organization={ACM}
}

@inproceedings{storey2014r,
  title={The (r) evolution of social media in software engineering},
  author={Storey, Margaret--Anne and Singer, Leif and Cleary, Brendan and Figueira Filho, Fernando and Zagalsky, Alexey},
  booktitle={Proceedings of the Future of Software Engineering},
  pages={100--116},
  year={2014},
  organization={ACM}
}

@article{dyba2005evidence,
  title={Evidence--based software engineering for practitioners},
  author={Dyb{\aa}, Tore and Kitchenham, Barbara A and J{\o}rgensen, Magne},
  journal={IEEE Software},
  volume={22},
  number={1},
  pages={58--65},
  year={2005},
  publisher={IEEE}
}

@article{rainer2017using,
  title={Using argumentation theory to analyse software practitioners' defeasible evidence, inference and belief},
  author={Rainer, Austen},
  journal={Information and Software Technology},
  year={2017},
  publisher={Elsevier}
}

@article{rainer2019usingblogs,
  title={Using blog--like documents to investigate software practice: Benefits, challenges, and research directions},
  author={Rainer, Austen and Williams, Ashley},
  journal={Journal of Software: Evolution and Process},
  volume={31},
  number={11},
  pages={e2197},
  year={2019},
  publisher={Wiley Online Library}
}

@article{soldani2018pains,
  title={The Pains and Gains of Microservices: A Systematic Grey Literature Review},
  author={Soldani, Jacopo and Tamburri, Damian Andrew and Van Den Heuvel, Willem--Jan},
  journal={Journal of Systems and Software},
  year={2018},
  publisher={Elsevier}
}

@article{rainer2020anaya,
  title={Anaya's Journey: A Vision for a Future Software Academy},
  author={Rainer, Austen},
  journal={IEEE Software},
  volume={37},
  number={2},
  pages={95--96},
  year={2020},
  publisher={IEEE}
}

@misc{kilimchoi18,
    author = {Kilim Choi},
    title = {Software Engineering Blogs},
    url  = {https://github.com/kilimchoi/engineering-blogs#-individuals},
    addendum = {(accessed: 04.08.2018)}
}

@misc{spolsky2006,
    author = {Joel Spolsky},
    title = {Language Wars},
    url  = {https://www.joelonsoftware.com/2006/09/01/language-wars/},
    addendum = {(accessed: 18.03.2020)}
}

@article{garousi2017guidelines,
  title={Guidelines for including the grey literature and conducting multivocal literature reviews in software engineering},
  author={Garousi, Vahid and Felderer, Michael and M{\"a}ntyl{\"a}, Mika V},
  journal={arXiv preprint arXiv:1707.02553},
  year={2017}
}

@inproceedings{garousi2016need,
  title={The need for multivocal literature reviews in software engineering: complementing systematic literature reviews with grey literature},
  author={Garousi, Vahid and Felderer, Michael and M{\"a}ntyl{\"a}, Mika V},
  booktitle={Proceedings of the 20th International Conference on Evaluation and Assessment in Software Engineering},
  pages={26},
  year={2016},
  organization={ACM}
}

@inproceedings{cabrio2018five,
  title={Five Years of Argument Mining: a Data-driven Analysis.},
  author={Cabrio, Elena and Villata, Serena},
  booktitle={IJCAI},
  pages={5427--5433},
  year={2018}
}

@article{lippi2016argumentation,
  title={Argumentation mining: State of the art and emerging trends},
  author={Lippi, Marco and Torroni, Paolo},
  journal={ACM Transactions on Internet Technology (TOIT)},
  volume={16},
  number={2},
  pages={10},
  year={2016},
  publisher={ACM}
}

@article{peldszus2013argument,
  title={From argument diagrams to argumentation mining in texts: A survey},
  author={Peldszus, Andreas and Stede, Manfred},
  journal={International Journal of Cognitive Informatics and Natural Intelligence (IJCINI)},
  volume={7},
  number={1},
  pages={1--31},
  year={2013},
  publisher={IGI Global}
}

@article{lippi2016margot,
  title={MARGOT: A web server for argumentation mining},
  author={Lippi, Marco and Torroni, Paolo},
  journal={Expert Systems with Applications},
  volume={65},
  pages={292--303},
  year={2016},
  publisher={Elsevier}
}

@article{mochales2011argumentation,
  title={Argumentation mining},
  author={Mochales, Raquel and Moens, Marie-Francine},
  journal={Artificial Intelligence and Law},
  volume={19},
  number={1},
  pages={1--22},
  year={2011},
  publisher={Springer}
}

@article{rainerIST18shortpaper,
  title={Heuristics for improving the rigour and relevance of grey literature searches for software engineering research},
  author={Rainer, Austen and Williams, Ashley},
  journal={Information and Software Technology},
  volume={106},
  pages={231--233},
  year={2019},
  publisher={Elsevier}
}

@article{bex2013values,
  title={Values as the point of a story},
  author={Bex, FJ},
  journal={From knowledge representation to argumentation in AI, law and policy making. A Festschrift in honour of Trevor Bench--Capon},
  pages={63--78},
  year={2013}
}

@book{twining1994rethinking,
  title={Rethinking evidence: Exploratory essays},
  author={Twining, William},
  year={1994},
  publisher={Northwestern University Press}
}

@Article{Tsafnat2014,
author={Tsafnat, Guy
and Glasziou, Paul
and Choong, Miew Keen
and Dunn, Adam
and Galgani, Filippo
and Coiera, Enrico},
title={Systematic review automation technologies},
journal={Systematic Reviews},
year={2014},
month={7},
day={09},
volume={3},
number={1},
pages={74},
issn={2046-4053},
doi={10.1186/2046-4053-3-74},
url={https://doi.org/10.1186/2046-4053-3-74}
}

@Article{OMara-Eves2015,
author="O'Mara--Eves, Alison
and Thomas, James
and McNaught, John
and Miwa, Makoto
and Ananiadou, Sophia",
title="Using text mining for study identification in systematic reviews: a systematic review of current approaches",
journal="Systematic Reviews",
year="2015",
month="1",
day="14",
volume="4",
number="1",
pages="5",
issn="2046-4053",
doi="10.1186/2046-4053-4-5",
url="https://doi.org/10.1186/2046-4053-4-5"
}

@inproceedings{marshall2014tools,
  title={Tools to support systematic reviews in software engineering: a feature analysis},
  author={Marshall, Christopher and Brereton, Pearl and Kitchenham, Barbara},
  booktitle={Proceedings of the 18th International Conference on Evaluation and Assessment in Software Engineering},
  pages={13},
  year={2014},
  organization={ACM}
}

@inproceedings{marshall2013tools,
  title={Tools to support systematic literature reviews in software engineering: A mapping study},
  author={Marshall, Christopher and Brereton, Pearl},
  booktitle={2013 ACM/IEEE International Symposium on Empirical Software Engineering and Measurement},
  pages={296--299},
  year={2013},
  organization={IEEE}
}

@inproceedings{singh2017exploring,
  title={Exploring Automatic Search in Digital Libraries: A Caution Guide for Systematic Reviewers},
  author={Singh, Paramvir and Singh, Karanpreet},
  booktitle={Proceedings of the 21st International Conference on Evaluation and Assessment in Software Engineering},
  pages={236--241},
  year={2017},
  organization={ACM}
}

@inproceedings{ceran2012hybrid,
  title={A hybrid model and memory based story classifier},
  author={Ceran, Betul and Karad, Ravi and Corman, Steven and Davulcu, Hasan},
  booktitle={Proceedings of the 3rd Workshop on Computational Models of Narrative},
  pages={58--62},
  year={2012}
}

@inproceedings{de2011collaborative,
  title={Collaborative narratives for business rule elicitation},
  author={de AR Gon{\c{c}}alves, Jo{\~a}o Carlos and Santoro, Fl{\'a}via Maria and Bai{\~a}o, Fernanda Araujo},
  booktitle={2011 IEEE International Conference on Systems, Man, and Cybernetics (SMC)},
  pages={1926--1931},
  year={2011},
  organization={IEEE}
}

@inproceedings{de2010case,
  title={A case study on designing business processes based on collaborative and mining approaches},
  author={de AR Gon{\c{c}}alves, Jo{\~a}o Carlos and Santoro, Fl{\'a}via Maria and Bai{\~a}o, Fernanda Araujo},
  booktitle={Computer Supported Cooperative Work in Design (CSCWD), 2010 14th International Conference on},
  pages={611--616},
  year={2010},
  organization={IEEE}
}

@article{garousi2016and,
  title={When and what to automate in software testing? A multi-vocal literature review},
  author={Garousi, Vahid and M{\"a}ntyl{\"a}, Mika V},
  journal={Information and Software Technology},
  volume={76},
  pages={92--117},
  year={2016},
  publisher={Elsevier}
}

@inproceedings{kurashima2006mining,
  title={Mining and visualizing local experiences from blog entries},
  author={Kurashima, Takeshi and Tezuka, Taro and Tanaka, Katsumi},
  booktitle={DEXA},
  pages={213--222},
  year={2006},
  organization={Springer}
}

@inproceedings{inui2008experience,
  title={Experience mining: Building a large-scale database of personal experiences and opinions from web documents},
  author={Inui, Kentaro and Abe, Shuya and Hara, Kazuo and Morita, Hiraku and Sao, Chitose and Eguchi, Megumi and Sumida, Asuka and Murakami, Koji and Matsuyoshi, Suguru},
  booktitle={Proceedings of the 2008 IEEE/WIC/ACM International Conference on Web Intelligence and Intelligent Agent Technology-Volume 01},
  pages={314--321},
  year={2008},
  organization={IEEE Computer Society}
}

@inproceedings{jijkoun2010mining,
  title={Mining user experiences from online forums: an exploration},
  author={Jijkoun, Valentin and de Rijke, Maarten and Weerkamp, Wouter and Ackermans, Paul and Geleijnse, Gijs},
  booktitle={Proceedings of the NAACL HLT 2010 Workshop on Computational Linguistics in a World of Social Media},
  pages={17--18},
  year={2010},
  organization={Association for Computational Linguistics}
}

@inproceedings{kuzey2014fresh,
  title={A fresh look on knowledge bases: Distilling named events from news},
  author={Kuzey, Erdal and Vreeken, Jilles and Weikum, Gerhard},
  booktitle={Proceedings of the 23rd ACM International Conference on Conference on Information and Knowledge Management},
  pages={1689--1698},
  year={2014},
  organization={ACM}
}

@inproceedings{brants2003system,
  title={A system for new event detection},
  author={Brants, Thorsten and Chen, Francine and Farahat, Ayman},
  booktitle={Proceedings of the 26th Annual International ACM SIGIR Conference on Research and Development in Information Retrieval},
  pages={330--337},
  year={2003},
  organization={ACM}
}

@inproceedings{nanni2017building,
  title={Building entity--centric event collections},
  author={Nanni, Federico and Ponzetto, Simone Paolo and Dietz, Laura},
  booktitle={Proceedings of the 17th ACM/IEEE Joint Conference on Digital Libraries},
  pages={199--208},
  year={2017},
  organization={IEEE Press}
}

@inproceedings{masterov2015canary,
  title={Canary in the e--commerce coal mine: Detecting and predicting poor experiences using buyer-to-seller messages},
  author={Masterov, Dimitriy V and Mayer, Uwe F and Tadelis, Steven},
  booktitle={Proceedings of the Sixteenth ACM Conference on Economics and Computation},
  pages={81--93},
  year={2015},
  organization={ACM}
}

@inproceedings{liu2015context,
  title={Context--aware experience extraction from online health forums},
  author={Liu, Yunzhong and Chen, Yi and Tang, Jiliang and Liu, Huan},
  booktitle={2015 International Conference on Healthcare Informatics (ICHI)},
  pages={42--47},
  year={2015},
  organization={IEEE}
}

@inproceedings{calix2017deep,
  title={Deep gramulator: Improving precision in the classification of personal health-experience tweets with deep learning},
  author={Calix, Ricardo A and Gupta, Ravish and Gupta, Matrika and Jiang, Keyuan},
  booktitle={2017 IEEE International Conference on Bioinformatics and Biomedicine (BIBM)},
  pages={1154--1159},
  year={2017},
  organization={IEEE}
}

@inproceedings{park2010detecting,
  title={Detecting experiences from weblogs},
  author={Park, Keun Chan and Jeong, Yoonjae and Myaeng, Sung Hyon},
  booktitle={Proceedings of the 48th Annual Meeting of the Association for Computational Linguistics},
  pages={1464--1472},
  year={2010},
  organization={Association for Computational Linguistics}
}

@inproceedings{kurashima2009discovering,
  title={Discovering association rules on experiences from large-scale blog entries},
  author={Kurashima, Takeshi and Fujimura, Ko and Okuda, Hidenori},
  booktitle={European Conference on Information Retrieval},
  pages={546--553},
  year={2009},
  organization={Springer}
}

@inproceedings{hassan2016event,
  title={Event--Based Recognition of Lived Experiences in User Reviews},
  author={Hassan, Ehab and Buscaldi, Davide and Gangemi, Aldo},
  booktitle={European Knowledge Acquisition Workshop},
  pages={320--336},
  year={2016},
  organization={Springer}
}

@inproceedings{garcia2010exploring,
  title={Exploring hotel service quality experience indicators in user-generated content: A case using Tripadvisor data},
  author={Garc{\'i}a-Barriocanal, Elena and Sicilia, Miguel-Angel and Korfiatis, Nikolaos},
  booktitle={Proceedings of the 5th Mediterranean Conference on Information Systems (MCIS 2010)},
  year={2010}
}

@inproceedings{hattori2012extracting,
  title={Extracting tip information from social media},
  author={Hattori, Yuki and Nadamoto, Akiyo},
  booktitle={Proceedings of the 14th International Conference on Information Integration and Web-based Applications \& Services},
  pages={205--212},
  year={2012},
  organization={ACM}
}

@inproceedings{fontao2017facing,
  title={Facing up the primary emotions in Mobile Software Ecosystems from Developer Experience},
  author={Font{\~a}o, Awdren and Ekwoge, Oswald M and Santos, Rodrigo and Dias-Neto, Arilo Claudio},
  booktitle={Proceedings of the 2nd Workshop on Social, Human, and Economic Aspects of Software},
  pages={5--11},
  year={2017},
  organization={ACM}
}

@inproceedings{wilson2014finding,
  title={Finding information about mental health in microblogging platforms: a case study of depression},
  author={Wilson, Max L and Ali, Susan and Valstar, Michel F},
  booktitle={Proceedings of the 5th Information Interaction in Context Symposium},
  pages={8--17},
  year={2014},
  organization={ACM}
}

@inproceedings{bonchi2016identifying,
  title={Identifying buzzing stories via anomalous temporal subgraph discovery},
  author={Bonchi, Francesco and Bordino, Ilaria and Gullo, Francesco and Stilo, Giovanni},
  booktitle={Web Intelligence (WI), 2016 IEEE/WIC/ACM International Conference on},
  pages={161--168},
  year={2016},
  organization={IEEE}
}

@article{min2012identifying,
  title={Identifying helpful reviews based on customer’s mentions about experiences},
  author={Min, Hye-Jin and Park, Jong C},
  journal={Expert Systems with Applications},
  volume={39},
  number={15},
  pages={11830--11838},
  year={2012},
  publisher={Elsevier}
}

@inproceedings{swanson2014identifying,
  title={Identifying narrative clause types in personal stories},
  author={Swanson, Reid and Rahimtoroghi, Elahe and Corcoran, Thomas and Walker, Marilyn},
  booktitle={Proceedings of the 15th Annual Meeting of the Special Interest Group on Discourse and Dialogue (SIGDIAL)},
  pages={171--180},
  year={2014}
}

@inproceedings{jiang2017identifying,
  title={Identifying personal health experience tweets with deep neural networks},
  author={Jiang, Keyuan and Gupta, Ravish and Gupta, Matrika and Calix, Ricardo A and Bernard, Gordon R},
  booktitle={39th IEEE Annual International Conference of Engineering in Medicine and Biology Society (EMBC)},
  pages={1174--1177},
  year={2017},
  organization={IEEE}
}

@article{lee2006korean,
  title={Korean--Japanese story link detection based on distributional and contrastive properties of event terms},
  author={Lee, Kyung-Soon and Kageura, Kyo},
  journal={Information processing \& management},
  volume={42},
  number={2},
  pages={538--550},
  year={2006},
  publisher={Elsevier}
}

@article{yu2018learning,
  title={Learning distributed sentence representations for story segmentation},
  author={Yu, Jia and Xie, Lei and Xiao, Xiong and Chng, Eng Siong},
  journal={Signal Processing},
  volume={142},
  pages={403--411},
  year={2018},
  publisher={Elsevier}
}

@inproceedings{qamra2006mining,
  title={Mining blog stories using community-based and temporal clustering},
  author={Qamra, Arun and Tseng, Belle and Chang, Edward Y},
  booktitle={Proceedings of the 15th ACM International Conference on Information and Knowledge Management},
  pages={58--67},
  year={2006},
  organization={ACM}
}

@inproceedings{gruenheid2015storypivot,
  title={StoryPivot: comparing and contrasting story evolution},
  author={Gruenheid, Anja and Kossmann, Donald and Rekatsinas, Theodoros and Srivastava, Divesh},
  booktitle={Proceedings of the 2015 ACM SIGMOD International Conference on Management of Data},
  pages={1415--1420},
  year={2015},
  organization={ACM}
}

@inproceedings{petrovic2010streaming,
  title={Streaming first story detection with application to twitter},
  author={Petrovi{\'c}, Sa{\v{s}}a and Osborne, Miles and Lavrenko, Victor},
  booktitle={Human language technologies: The 2010 Annual Conference of the North American Chapter of the Association for Computational Linguistics},
  pages={181--189},
  year={2010},
  organization={Association for Computational Linguistics}
}

@article{srijith2017sub,
  title={Sub-story detection in Twitter with hierarchical Dirichlet processes},
  author={Srijith, PK and Hepple, Mark and Bontcheva, Kalina and Preotiuc-Pietro, Daniel},
  journal={Information Processing \& Management},
  volume={53},
  number={4},
  pages={989--1003},
  year={2017},
  publisher={Elsevier}
}

@article{rahimtoroghi2013evaluation,
  title={Evaluation, orientation, and action in interactive storytelling},
  author={Rahimtoroghi, Elahe and Swanson, Reid and Walker, Marilyn A and Corcoran, Thomas},
  journal={Proceedings of Intelligent Narrative Technologies},
  volume={6},
  year={2013}
}

@inproceedings{gordon2009identifying,
  title={Identifying personal stories in millions of weblog entries},
  author={Gordon, Andrew and Swanson, Reid},
  booktitle={Third International Conference on Weblogs and Social Media, Data Challenge Workshop, San Jose, CA},
  volume={46},
  year={2009}
}

@inproceedings{parnin2011measuring,
  title={Measuring API documentation on the web},
  author={Parnin, Chris and Treude, Christoph},
  booktitle={Proceedings of the 2nd international Workshop on Web 2.0 for Software Engineering},
  pages={25--30},
  year={2011},
  organization={ACM}
}

@incollection{labov1967narrative,
  title={Narrative analysis. Essays on the verbal and visual arts},
  author={Labov, William and Waletzky, Joshua},
  booktitle={Essays on the verbal and visual arts},
  editor={Helm, June and MacNeish, June Helm},
  year={1967},
  publisher={University of Washington Press, Seattle}
}

@article{achour1999guiding,
  title={Guiding scenario authoring},
  author={Achour, Camille Ben},
  journal={Information Modelling and Knowledge Bases X},
  volume={51},
  pages={152},
  year={1999},
  publisher={IOS Press}
}

@book{vendler1967linguistics,
  title={Linguistics in Philosophy. ithaca},
  author={Vendler, Zeno},
  journal={NY: Cornell UP},
  year={1967}
}

@inproceedings{ceran2012asemantic,
 author = {Ceran, Betul and Karad, Ravi and Mandvekar, Ajay and Corman, Steven R. and Davulcu, Hasan},
 title = {A Semantic Triplet Based Story Classifier},
 booktitle = {Proceedings of the 2012 International Conference on Advances in Social Networks Analysis and Mining (ASONAM 2012)},
 year = {2012},
 isbn = {978-0-7695-4799-2},
 pages = {573--580},
 numpages = {8},
 url = {http://dx.doi.org/10.1109/ASONAM.2012.97},
 doi = {10.1109/ASONAM.2012.97},
 acmid = {2457117},
 publisher = {IEEE Computer Society},
 address = {Washington, DC, USA},
}

@article{goswami2016survey,
author="Goswami, Anuradha and Kumar, Ajey",
title="A survey of event detection techniques in online social networks",
journal="Social Network Analysis and Mining",
year="2016",
month="11",
day="17",
volume="6",
number="1",
pages="107",
doi="10.1007/s13278-016-0414-1",
url="https://doi.org/10.1007/s13278-016-0414-1"
}

@inproceedings{DiCrescenzo2017hermevent,
 author = {Di Crescenzo, Cristiano and Gavazzi, Giulia and Legnaro, Giacomo and Troccoli, Elena and Bordino, Ilaria and Gullo, Francesco},
 title = {HERMEVENT: A News Collection for Emerging-event Detection},
 booktitle = {Proceedings of the 7th International Conference on Web Intelligence, Mining and Semantics},
 series = {WIMS '17},
 year = {2017},
 isbn = {978-1-4503-5225-3},
 location = {Amantea, Italy},
 pages = {11:1--11:10},
 articleno = {11},
 numpages = {10},
 url = {http://doi.acm.org/10.1145/3102254.3102262},
 doi = {10.1145/3102254.3102262},
 acmid = {3102262},
 publisher = {ACM},
 address = {New York, NY, USA},
}

@article{khrouf2014mining,
  title={Mining events connections on the social web: Real-time instance matching and data analysis in EventMedia},
  author={Khrouf, Houda and Milicic, Vuk and Troncy, Rapha{\"e}l},
  journal={Web Semantics: Science, Services and Agents on the World Wide Web},
  volume={24},
  pages={3--10},
  year={2014},
  publisher={Elsevier}
}

@article{abe2011mining,
  title={Mining personal experiences and opinions from Web documents},
  author={Abe, Shuya and Inui, Kentaro and Hara, Kazuo and Morita, Hiraku and Sao, Chitose and Eguchi, Megumi and Sumita, Asuka and Murakami, Koji and Matsuyoshi, Suguru},
  journal={Web Intelligence and Agent Systems: An International Journal},
  volume={9},
  number={2},
  pages={109--121},
  year={2011},
  publisher={IOS Press}
}

@inproceedings{lee2004multilingual,
  title={Multilingual story link detection based on event term weighting on times and multilingual spaces},
  author={Lee, Kyung-Soon and Kageura, Kyo},
  booktitle={International Conference on Asian Digital Libraries},
  pages={398--407},
  year={2004},
  organization={Springer}
}

@article{krawczyk2016perceptual,
  title={Perceptual mapping of hotel brands using online reviews: a text analytics approach},
  author={Krawczyk, Matthew and Xiang, Zheng},
  journal={Information Technology \& Tourism},
  volume={16},
  number={1},
  pages={23--43},
  year={2016},
  publisher={Springer}
}

@inproceedings{mazoyer2018realtime,
  title={Real-time collection of reliable and representative tweets datasets related to news events},
  author={B\'eatrice Mazoyer and Julia Cag\'e and C\'eline Hudelot and Marie--Luce Viaud1},
  booktitle={BroDyn 2018: 1st Workshop on Analysis of Broad Dynamic Topics over Social Media},
  year={2018}
}

@ARTICLE{Behrooz2015remember,
author={Behrooz, M. and Swanson, R. and Jhala, A.},
title={Remember that time? Telling interesting stories from past interactions},
journal={Lecture Notes in Computer Science (including subseries Lecture Notes in Artificial Intelligence and Lecture Notes in Bioinformatics)},
year={2015},
volume={9445},
pages={93-104},
doi={10.1007/978-3-319-27036-4_9},
note={cited By 1},
url={https://www.scopus.com/inward/record.uri?eid=2-s2.0-84951871183&doi=10.1007\%2f978-3-319-27036-4_9&partnerID=40&md5=ce1b5a85586a8a12ee9d78efdd93fa48},
document_type={Conference Paper},
source={Scopus},
}

@INPROCEEDINGS{ceran2015story,
author={B. Ceran and N. Kedia and S. R. Corman and H. Davulcu},
booktitle={2015 IEEE/ACM International Conference on Advances in Social Networks Analysis and Mining (ASONAM)},
title={Story detection using generalized concepts and relations},
year={2015},
volume={},
number={},
pages={942-949},
doi={10.1145/2808797.2809312},
ISSN={},
month={8},}

@article{hattori2013tip,
author = {Yuki Hattori and Akiyo Nadamoto},
title = {Tip information from social media based on topic detection},
journal = {International Journal of Web Information Systems},
volume = {9},
number = {1},
pages = {83-94},
year = {2013},
doi = {10.1108/17440081311316406},

URL = {
        https://doi.org/10.1108/17440081311316406
},
eprint = {https://doi.org/10.1108/17440081311316406
}, 
}

@CONFERENCE{Vails-Vargas2017towards,
author={Vails-Vargas, J. and Zhu, J. and Ontanon, S.},
title={Towards automatically extracting story graphs from natural language stories},
journal={AAAI Workshop - Technical Report},
year={2017},
volume={WS-17-01 - WS-17-15},
pages={1006-1013},
note={cited By 0},
url={https://www.scopus.com/inward/record.uri?eid=2-s2.0-85046093315&partnerID=40&md5=4f5906d4e39d5b384381758d698a589e},
document_type={Conference Paper},
source={Scopus},
}

@article{noferesti2015using,
title = "Using Linked Data for polarity classification of patients’ experiences",
journal = "Journal of Biomedical Informatics",
volume = "57",
pages = "6 - 19",
year = "2015",
issn = "1532-0464",
doi = "https://doi.org/10.1016/j.jbi.2015.06.017",
url = "http://www.sciencedirect.com/science/article/pii/S1532046415001276",
author = "Samira Noferesti and Mehrnoush Shamsfard",
}

@inproceedings{petrovic2012using,
 author = {Petrovi\'{c}, Sa\v{s}a and Osborne, Miles and Lavrenko, Victor},
 title = {Using Paraphrases for Improving First Story Detection in News and Twitter},
 booktitle = {Proceedings of the 2012 Conference of the North American Chapter of the Association for Computational Linguistics: Human Language Technologies},
 series = {NAACL HLT '12},
 year = {2012},
 isbn = {978-1-937284-20-6},
 location = {Montreal, Canada},
 pages = {338--346},
 numpages = {9},
 url = {http://dl.acm.org/citation.cfm?id=2382029.2382072},
 acmid = {2382072},
 publisher = {Association for Computational Linguistics},
 address = {Stroudsburg, PA, USA},
}

@article{garousi2018guidelines,
  title={Guidelines for including grey literature and conducting multivocal literature reviews in software engineering},
  author={Garousi, Vahid and Felderer, Michael and M{\"a}ntyl{\"a}, Mika V},
  journal={Information and Software Technology},
  year={2018},
  publisher={Elsevier}
}

@article{domingos2012few,
  title={A few useful things to know about machine learning},
  author={Domingos, Pedro},
  journal={Communications of the ACM},
  volume={55},
  number={10},
  pages={78--87},
  year={2012},
  publisher={ACM}
}

@inproceedings{burton2009icwsm,
  title={The ICWSM 2009 Spinn3r dataset},
  author={Burton, Kevin and Java, Akshay and Soboroff, Ian and others},
  booktitle={Third Annual Conference on Weblogs and Social Media (ICWSM 2009)},
  year={2009}
}

@inproceedings{ott2011finding,
  title={Finding deceptive opinion spam by any stretch of the imagination},
  author={Ott, Myle and Choi, Yejin and Cardie, Claire and Hancock, Jeffrey T},
  booktitle={Proceedings of the 49th Annual Meeting of the Association for Computational Linguistics: Human Language Technologies-Volume 1},
  pages={309--319},
  year={2011},
  organization={Association for Computational Linguistics}
}

@article{hemmatian2019survey,
  title={A survey on classification techniques for opinion mining and sentiment analysis},
  author={Hemmatian, Fatemeh and Sohrabi, Mohammad Karim},
  journal={Artificial Intelligence Review},
  pages={1--51},
  year={2019},
  publisher={Springer}
}

@article{chen2019bibliometric,
  title={A bibliometric analysis of event detection in social media},
  author={Chen, Xieling and Wang, Shan and Tang, Yong and Hao, Tianyong},
  journal={Online Information Review},
  year={2019},
  publisher={Emerald Publishing Limited}
}

@article{adams2016searching,
  title={Searching and synthesising ‘grey literature’and ‘grey information’in public health: critical reflections on three case studies},
  author={Adams, Jean and Hillier-Brown, Frances C and Moore, Helen J and Lake, Amelia A and Araujo-Soares, Vera and White, Martin and Summerbell, Carolyn},
  journal={Systematic reviews},
  volume={5},
  number={1},
  pages={164},
  year={2016},
  publisher={BioMed Central}
}

% \bibliography{bibliography}

\end{document}